\documentclass[aps,prl,twocolumn,floatfix,longbibliography]{revtex4-1}
\usepackage{bm}
\usepackage{epsf}
\usepackage{amssymb}
\usepackage{amsmath}
\usepackage{graphicx}
\usepackage{rotating}
\usepackage{epsfig}
\usepackage{psfrag}
\usepackage{amsmath}
\usepackage{hyperref}
\usepackage{subfigure}
\usepackage{braket}
\usepackage{tikz}
\usepackage{stmaryrd}
\usepackage{wasysym}

\newcommand{\bst}{\mathcal{T}}
\newcommand{\bea}{\begin{eqnarray}}
\newcommand{\eea}{\end{eqnarray}}
\newcommand{\bpm}{\begin{pmatrix}}
\newcommand{\epm}{\end{pmatrix}}

\newcommand{\imth}{\hspace{1pt}\mathrm{i}\hspace{1pt}}
\newcommand{\expval}[1]{\langle{#1}\rangle}

\begin{document}
\title{Magnetic fragmentation and fractionalized Goldstone modes in a bilayer quantum spin liquid}

\author{Aayush Vijayvargia$^1$, Emilian Marius Nica$^{1,2}$, Roderich Moessner$^3$, Yuan-Ming Lu$^{4}$, Onur Erten$^1$}
\affiliation{$^1$Department of Physics, Arizona State University, Tempe, AZ 85287, USA \\ $^{2}$Department of Physics and Astronomy, Rice University, 6100 Main St, Houston 77005 TX, USA \\ $^3$Max-Planck-Institut f\"ur Physik komplexer Systeme, N\"othnitzer Strasse 38, 01187 Dresden, Germany\\ $^4$ Department of Physics, The Ohio State University, Columbus OH 43210, USA}

\begin{abstract}
We study the phase diagram of a bilayer quantum spin liquid model with Kitaev-type interactions on a square lattice. We show that the low energy limit is described by a $\pi$-flux Hubbard model with an enhanced SO(4) symmetry. The antiferromagnetic Mott transition of the Hubbard model signals a magnetic fragmentation transition for the spin and orbital degrees of freedom of the bilayer. The fragmented ``N\'eel order'' features a non-local string order parameter for an in-plane N\'eel component, in addition to an anisotropic local order parameter. The associated quantum order is characterized by an emergent $\mathbb{Z}_{2} \times \mathbb{Z}_{2}$ gauge field when the N\'eel vector is along the $\hat{z}$ direction, and a $\mathbb{Z}_2$ gauge field otherwise. We underpin these results with a perturbative calculation, which is consistent with the field theory analysis. We conclude with a discussion on the low energy collective excitations of these phases and show that the Goldstone boson of the $\mathbb{Z}_{2} \times \mathbb{Z}_{2}$ phase is fractionalized and non-local. 
\end{abstract}
\maketitle
Quantum spin liquids (QSLs) are frustrated magnets that do not exhibit long range magnetic order down to zero temperature\cite{Broholm_Science2020, Balents_Nature2010, Savary_RepProgPhys2016,moessner_moore_2021}. Quantum fluctuations in these systems give rise to exotic phenomena such as fractionalization and long-range entanglement, which now become the defining properties for QSLs\cite{Zhou_RMP2017, Knolle_AnnRevCondMatPhys2019, Wen_RMP2017}. The Kitaev model on the honeycomb lattice\cite{Kitaev_AnnPhys2006} is one of the 
few examples of an exactly solvable model with a QSL ground state (GS). In recent years, remarkable progress in identifying candidate materials with strong Kitaev-type interactions has been achieved, in such instances as the iridates\cite{HwanChun_NatPhys2015, Kitagawa_Nat2018} and $\alpha$-RuCl$_3$ \cite{Takagi_NatRevPhys2019}. Kitaev interactions may also be strong in other van der Waals (vdW) materials\cite{Lee_PRL2020}. Bilayers and moir\'e superlattices of vdW materials are new tunable quantum platforms for realizing a multitude of novel phases, with a variety of basic building blocks including graphene\cite{Cao2018}, semiconductors\cite{Devakul_NatComm2021} and superconductors\cite{Zhao_arxiv2021}. 
\begin{figure}[t]
\includegraphics[width=7.9cm]{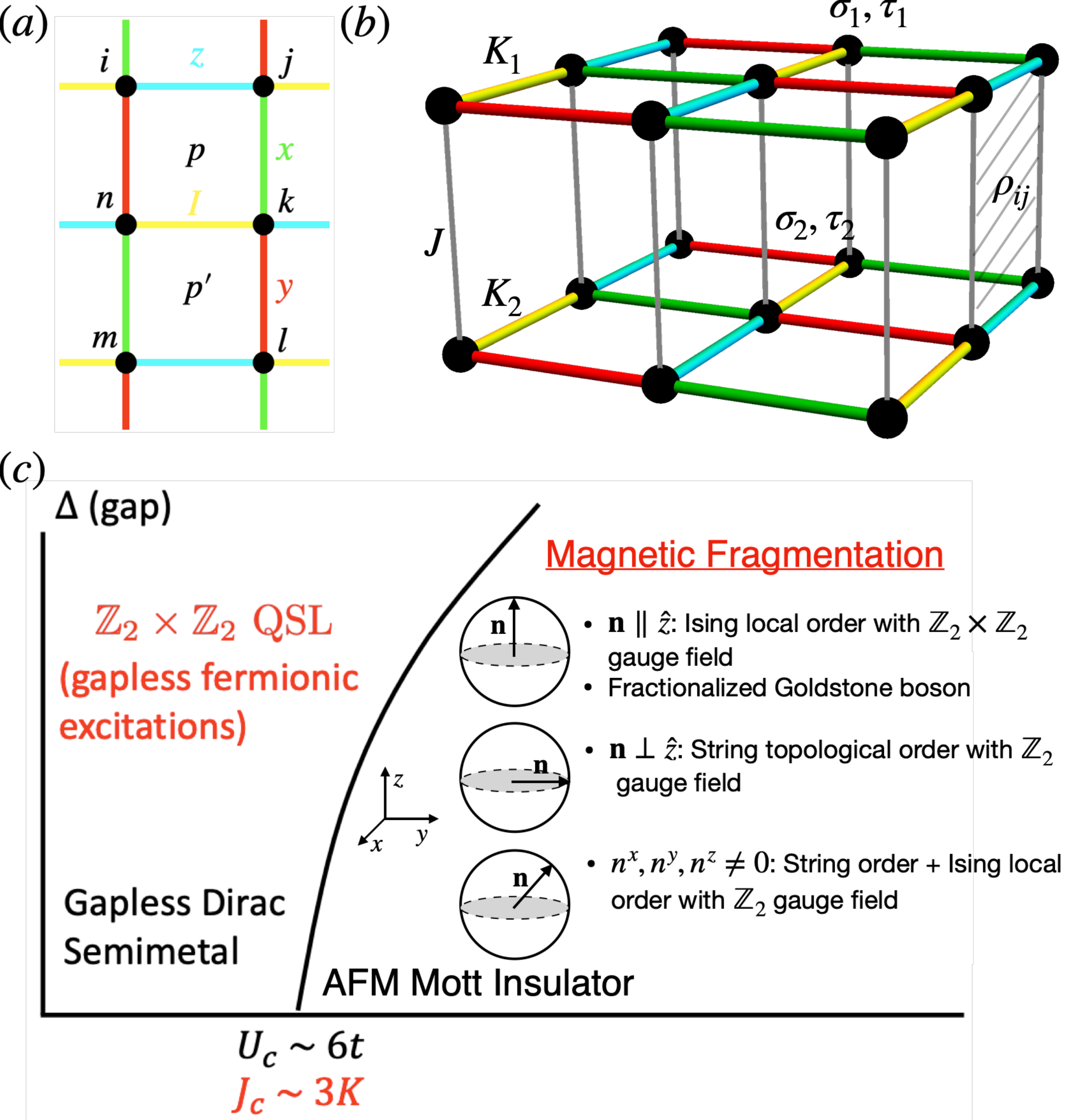}
\caption{Schematic of the model and the phase diagram: (a) single layer unit cell. Four different colors depict four types of bonds. There are two inequivalent plaquettes $p$ and $p'$ in a unit cell. (b) Bilayer model with intra-layer Kitaev parameters $K_\nu$ and an inter-layer exchange, $J$. (c) The low-energy description of the model is a $\pi$-flux Hubbard model which exhibits a Mott transition at $U/t \sim 6$ (black). In terms of the original degrees of freedom, the Mott transition corresponds to a magnetic fragmentation transition where a local magnetic order coexists with a non-local topological order.}
\label{Fig:1}
\end{figure}

Motivated by these developments, we study the phase diagram of a bilayer QSL model with Kitaev-type interactions on a square lattice (see Fig~\ref{Fig:1}(a)). First introduced in Ref. \citenum{Nakai_PRB2012}, the exact ground state of the monolayer model is an algebraic QSL featuring two flavors of Majorana fermions that are delocalized on the $\pi$-flux square lattice, and gapped $\pi$-flux (vison) excitations\cite{Seifert_PRL2020}. In the bilayer model Eq. (\ref{eq:H}), we add an Ising-type interlayer spin interaction, which commutes with the intra-layer flux operators and hence allows for controlled calculations. Our main results are summarized as follows: (i) Below the vison gap, we map the low-energy subspace of the bilayer model to a $\pi$-flux Hubbard model at half-filling, with an emergent $SO(4)$ symmetry. Monte Carlo studies of this model show an antiferromagnetic (AFM) Mott transition at critical $U_c \sim 6t$ \cite{Otsuka2013}. (ii) The in-plane components of the AFM order parameter, i.e. $n^{x,y}$ of the N\'eel vector ${\bf n}$, correspond to non-local order parameters whereas the out-of-plane component ($n^z$) is a local order parameter in terms of the spin and orbital degrees of freedom (DOF) of the bilayer system. (iii) The system features a $\mathbb{Z}_2 \times \mathbb{Z}_2$ gauge field when the N\'eel vector points along $\hat{z}$, and a $\mathbb{Z}_2$ gauge field otherwise. (iv) To complement the results of the Hubbard model, we perturbatively derive an effective Hamiltonian in the limit of large interlayer interactions. We confirm the magnetic fragmentation and topological degeneracy directly in terms of the original DOF, which are consistent with the Majorana fermion representation of the spin model. (v) We show that the Goldstone modes of the fragmented AFM order is fractionalized in the $\mathbb{Z}_2 \times \mathbb{Z}_2$ phase, in comparison to the normal Goldstone modes in the $\mathbb{Z}_2$ phase. 

\noindent
{\bf Microscopic model.}
One of the key conditions for the exact solution of the Kitaev model is the anticommutation relations of the Pauli matrices, $\{\sigma_{i}, \sigma_j \} = 2\delta_{ij}$. Since there are only three Pauli matrices, this method can only be applied to lattices with coordination number $z=3$ such as honeycomb, hyperhoneycomb and hyperoctagon lattices. However, it is possible to extend Kitaev's method to $\Gamma$ matrices that obey the Clifford algebra $\{\Gamma_{i}, \Gamma_j \} = 2\delta_{ij}$ \cite{Wu_PRB2009,Yao_PRL2009}. For instance, for a four-dimensional representation of the Clifford algebra, there are five $\Gamma^{\alpha}$ operators along with ten $\Gamma^{\alpha \beta}=\frac{i}{2}[\Gamma^{\alpha},\Gamma^{\beta}]$ and an identity matrix, which span the local Hilbert space. Therefore, Kitaev's construction can be extended to lattices with coordination number up to $z=5$ \cite{Wu_PRB2009,Yao_PRL2009}. We adapt this representation and consider the intra-layer Hamiltonian \cite{Nakai_PRB2012}, $H_K=-\sum_{\langle ij \rangle_{\gamma},\nu}K_\nu(\Gamma_{\nu i}^{\gamma}\Gamma_{\nu j}^{\gamma}+\Gamma_{\nu i}^{\gamma5}\Gamma_{\nu j}^{\gamma5})$, where $\nu=1,2$ is the layer index and $\gamma$ is the type of the bond as depicted in Fig. \ref{Fig:1}.  We also introduce an inter-layer Ising interaction $H_J= J\sum_i\Gamma^5_{1i}\Gamma^5_{2i}$. The full Hamiltonian can be expressed in terms of spins $(\sigma)$ and orbital $(\tau)$ Pauli matrices using the relation $\Gamma^{\alpha}=-\sigma^y\otimes \tau^{\alpha} \quad (\alpha=x,y,z)$, $\Gamma^4=\sigma^x\otimes\mathbb{I}_2$ and $\Gamma^5=-\sigma^z\otimes\mathbb{I}_2$,
\begin{align}
  \nonumber  &H=H_K+H_J=\\
    &-\sum_{\langle ij \rangle_\gamma,\nu }K_\nu(\sigma^x_{\nu i}\sigma^x_{\nu j} +\sigma^y_{\nu i}\sigma^y_{\nu j})(\tau_{\nu i}^{\gamma}\tau_{\nu j}^{\gamma})+J\sum_i\sigma^{z}_{1i}\sigma^{z}_{2i}
    \label{eq:H}
\end{align}
Here $\tau^{\gamma} = \tau^x,\tau^y,\tau^z,\mathbb{I}$ for $\gamma=1,2,3,4$ respectively, corresponding to the 4 bonds incident on a vertex of the square lattice as shown in Fig. \ref{Fig:1}(a) and the sum is over all the $\gamma$ bonds. Note that the $\gamma = 4$ (yellow) bond, which we refer as the `identity' bond henceforth, has trivial orbital dependence. We consider $K_1=K_2=K$, unless specified otherwise. We identify two inequivalent intra-layer flux plaquette operators $W_{\nu p}=\sigma^z_{\nu k}\sigma^z_{\nu n}\tau^x_{\nu i}\tau^y_{\nu j}\tau^x_{\nu k}\tau^y_{\nu n}$ and $W_{\nu p^\prime}=\sigma^z_{\nu k}\sigma^z_{\nu n}\tau^x_{\nu n}\tau^y_{\nu k}\tau^x_{\nu l}\tau^y_{\nu m}$, each with $\pm$ 1 eigenvalues. Both types of
plaquette operators commute with the Hamiltonian and the Hilbert space is divided into sectors of conserved fluxes. Note that the Ising form of the inter-layer exchange is crucial to preserve $[W_{\nu p/p^{\prime}},H]=0$ \cite{Seifert_PRL2020}. The intra-layer Hamiltonian can be solved by using a Majorana fermion representation of the $\Gamma$ matrices\cite{Nakai_PRB2012}, $H_K=K\sum_{\langle ij \rangle_\gamma,\nu }iu^\gamma_{\nu, ij}[c^x_{\nu i}c^x_{\nu j}+c^y_{\nu i}c^y_{\nu j}]$ where $u^\gamma_{\nu, ij}=ib^{\gamma}_{\nu i}b^{\gamma}_{\nu j}$ (see supplemental material (SM) for details). This representation is redundant and the physical states in each layer must be restricted to the eigenstates of $D_{\nu j}=ib^1_{\nu j}b^2_{\nu j}b^3_{\nu j}b^4_{\nu j}c^x_{\nu j}c^y_{\nu j}$, with eigenvalues 1. As in the Kitaev model, these constraints are imposed by the projection operator $P_\nu=\prod_i(1+D_{\nu i})/2$. The intra-layer bond operators, $u^\gamma_{\nu, ij}$ commute with $H_K$ and therefore are conserved with eigenvalues $\pm 1$. A $\mathbb{Z}_{2}$ gauge transformation at site $i$ for layer $\nu$ involves flipping the signs of the Majorana fermions and bond operators, $c_{\nu i}^{\alpha} \rightarrow -c_{\nu i}^{\alpha};~ u^\gamma_{\nu, \braket{ij}} \rightarrow -u^\gamma_{\nu, \braket{ij}}$. 
 %\en{excluding the trivial identity bonds}.
 %Two flavors of itinerant Majorana fermions are coupled to a $\mathbb{Z}_{2}$ gauge field described by the vison fields on each link described by $u_{ij}$. 
We combine the Majorana fermions on the two layers to form complex fermions, $f_{\nu i}=(c^x_{\nu i}-ic^y_{\nu i})/2$ such that
$ H_K=2K\sum_{\langle ij \rangle}u^\gamma_{\nu,ij}[if_{\nu i}^{\dag}f_{\nu j}+ {\rm H.c.}]$. 

According to Lieb's theorem \cite{Lieb_PRL1994}, the GS manifold of $H_K$ lies in the $\pi$-flux sector and consequently the eigenvalue of $W_{\nu p/p^{\prime}} = \prod_{p/p^{\prime}} u^\gamma_{\nu ij}$ is $-1$ in any GS configuration, for all square plaquettes. 
The spectrum is given by $E_K=\pm 4K\sqrt{\cos^2{k_x}+\sin^2{k_y}}$ which includes two inequivalent Dirac points at $(\pm \frac{\pi}{2},0)$.

Next, we represent the inter-layer interaction in terms of the Majorana fermions: $H_J=-J\sum_ic^x_{1i}c^y_{1i}c^x_{2i}c^y_{2i}$. $H_J$ commutes with the intra-layer flux operators $W_{p/p{\prime}}$. However, the quartic form of the inter-layer exchange precludes the exact solvability of $H$, which can be expressed as
\begin{eqnarray}
      H&=&2K\sum_{\langle ij \rangle_\gamma,\nu }u^\gamma_{\nu,ij}[if_{\nu i}^{\dag}f_{\nu j}+ \rm H.c.] \nonumber \\
      &+& 2J\sum_i[n_{1 i}+n_{2i}-1]^2
      \label{eq:Hcf}
\end{eqnarray}
where $n_{\nu i}=f^\dag_{\nu i}f_{\nu i}$.

\noindent
{\bf Enhanced emergent symmetry.} 
The Hamiltonian in eq. \ref{eq:Hcf} has a global $U(1)$ symmetry in each layer ($\nu=1,2$), $e^{-i\theta\sum_{i} \sigma^z_{\nu i}}H e^{i\theta\sum_{i}\sigma^z_{\nu i}}=H$, a $\mathbb{Z}_2$ layer exchange symmetry $\mathcal{X}$, a particle-hole symmetry $\mathcal{C}$, and a time reversal symmetry $\bst$. This results in a full symmetry group $G=O(2)_c\times O(2)_s\times \mathbb{Z}_2^\bst$ of model (\ref{eq:H}), as detailed in SM. After the Majoranization, the two $U(1)$ rotations manifest themselves as:
\begin{align}
    \nonumber &U_c(\theta) f_{\nu i}U_c^{-1}(\theta)=e^{-i \theta}f_{\nu i},\\
   & U_s(\theta) f_{\nu i}U_s^{-1}(\theta)=e^{-i \kappa \theta}f_{\nu i}
\end{align}
where $\kappa=-1,1$ for $\nu=1,2$ respectively.
This can be viewed as ``charge'': $U_c=e^{i\theta\sum_{\nu i}f^\dag_{\nu i}f_{\nu i}}$ and ``pseudo-spin'': $U_s=e^{i\theta\sum_{\nu i}\kappa f^\dag_{\nu i}f_{\nu i}}$ rotations where $\kappa = +1(-1)$ for $\nu = 1(2)$. The particle-hole symmetry $\mathcal{C}$ and $U(1)$ charge rotations form the $O(2)_c$ subgroup, while the layer exchange $\mathcal{X}$ and $U(1)$ pseudo-spin rotations form the $O(2)_s$ subgroup.  
% $f_{i,\kappa}\rightarrow f_{i,-\kappa}; ~ b^\gamma_{i,\kappa}\rightarrow b^\gamma_{i,-\kappa}$.
%\begin{align}
%f_{i,\kappa}\overset{\hat X}\longrightarrow f_{i,-\kappa}, \quad  b^\gamma_{i,\kappa}\overset{\hat X}\longrightarrow b^\gamma_{i,-\kappa}
%\end{align}
Next, we fix the gauge by choosing $u^\gamma_{\nu, ij}=u_{ij}$, for both the layers and pick the $\pi$-flux configuration as discussed above.
%since the Hilbert space is divided into sub-sectors with fixed gauge configurations, we can choose a gauge configuration $u_{ij}^\nu=u_{ij}$, for both the layers. After fixing the gauge, 
The resulting low-energy Hamiltonian (below the flux/vison gap) is a $\pi$-flux Hubbard model at half-filling with a hopping amplitude $t=2K$ and interaction strength $U=4J$.
%, corresponding to the fermions on the two layers, hopping on a square lattice with a background flux. The inter-layer interaction term is nothing but the on-site Hubbard interaction term.As was discussed earlier, the ground state flux at each plaqutte is $\pi$, as predicted by Lieb's theorem. The ground state hence, given its fixed flux configuration, is identical to the Hubbard model. 
It is well established \cite{Yang1990} that the Hubbard model on a bipartite lattice possesses an enhanced $G^\prime=SO(4)\times \mathbb{Z}_2^\bst=\mathbb{Z}_2^\bst\times SU(2)_c\times SU(2)_s/\mathbb{Z}_2$ symmetry. The equivalence established above shows that our model also exhibits an enhanced $SO(4)$ symmetry at the low energy sector. In fact, this emergent $SO(4)$ symmetry exists in any subspace with a fixed flux configuration. Emergent symmetries can play a key role to describe the low energy physics of strongly correlated systems including cuprates\cite{Demler_RMP2004} and iron pnictides\cite{Podolsky_2009}.

Quantum Monte Carlo studies have shown that the repulsive ($J>0$) $\pi$-flux Hubbard model displays a phase transition from Dirac semimetal to an AFM Mott insulator at $J/K \sim 3$ ($U/t \sim 6$) \cite{Chang2012} (see Fig. \ref{Fig:1}(c)). Moreover, due to $J \rightarrow -J$ mapping in the Hubbard model, the phase diagram is symmetric for ferromagnetic (FM) and AFM inter-layer exchange, for which the N\'eel order maps to superconducting and charge density wave orders.  

The N\'eel vector of the AFM order
\begin{eqnarray}\label{Neel vector}
\textbf{n} =\frac{1}{N}\sum_i{\bf n}_i,~~~{\bf n}_i = (-1)^{r_{ix}+r_{iy}}\langle f^\dagger_{\mu i}\boldsymbol \sigma_{\mu \nu}f_{\nu i}\rangle,
\end{eqnarray}
where $N$ is the number of sites and $r_{ix(y)}$ is the $x(y)$ coordinate of site $i$, can point along any direction on the Bloch sphere. Goldstone modes always arise since the emergent symmetry $G^\prime=SO(4)\times \mathbb{Z}_2^\bst$ is spontaneously broken down to $H^\prime=SU(2)_c\times U(1)_s\rtimes \mathbb{Z}_2^{\tilde\bst}$ in the N\'eel order (see SM for details). However, as we show below, different orientations of the N\'eel vector correspond to distinct ground states with different symmetry and topological properties, as summarized in Table \ref{tab:phases}.

\begin{table}[t]
\begin{tabular}{|c|c|c|c|}
\hline
N\'eel vector&Unbroken subgroup $H$&$G/H$&Gauge group\\
\hline
${\bf n}\parallel\hat z$&$O(2)_c\times U(1)_s\rtimes \mathbb{Z}_2^{\mathcal{X}\cdot\bst}$&$\mathbb{Z}_2$&$\mathbb{Z}_2\times\mathbb{Z}_2$\\
\hline
${\bf n}\perp\hat z$&$O(2)_c\times \mathbb{Z}_2\times \mathbb{Z}_2^{\tilde\bst}$&$S^1$&$\mathbb{Z}_2$\\
\hline
$n_z\neq0,n_xn_y\neq0$&$O(2)_c\times \mathbb{Z}_2^{\bst_{\bf n}}$&$O(2)$&$\mathbb{Z}_2$\\
\hline
\end{tabular}
\caption{Distinct ground state phases associated with different orientations of the N\'eel vector ${\bf n}$ in (\ref{Neel vector}). In each phase, the full symmetry $G=O(2)_c\times O(2)_s\times \mathbb{Z}_2^\bst$ of model (\ref{eq:H}) is spontaneously broken down to a different subgroup $H\leq G$, with an order parameter manifold $\mathcal{M}=G/H$. The gauge group for the associated topological order in each phase is also listed.}\label{tab:phases}
\end{table}

\noindent
{\it (i)} The N\'eel vector points along the z-direction, ${\bf n} \parallel \hat{z}$,
\begin{align}
    n_i^z = (-1)^{r_{ix}+r_{iy}}\expval{f^\dagger_{1i}f_{1i}-f^\dagger_{2i}f_{2i}}\neq0
    \label{eq:Nz}
\end{align}
In terms of Majorana fermions, eq.~\ref{eq:Nz} takes the form $n_i^z =(-1)^{r_{ix}+r_{iy}} i(c^x_{1i}c^y_{1i}-c^x_{2i}c^y_{2i})$. Note that $n_i^z$ is invariant under local $\mathbb{Z}_2$ gauge transformations $(c_{\nu i}^x, c_{\nu i}^y) \rightarrow (-c_{\nu i}^x, -c_{\nu i}^y)$, hence corresponding to a physical operator $(-1)^{r_{ix}+r_{iy}}(\sigma_{1i}^z-\sigma_{2i}^z)$. In other words, $n^z$ is a local order parameter of a Landau-type long range order.

%In this case, the z-component of the N\'eel vector, which can be rewritten as  $\frac1V\sum_i(-1)^i\expval{\sigma_{1i}^z-\sigma_{2i}^z}$, which makes the N\'eel spin order more evident. This is a gauge invariant quantity, which can be noted by looking at the Majorana fermion form: $\frac1V\sum_i(-1)^i\expval{i(c^x_{1i}c^y_{1i}-c^x_{2i}c^y_{2i})}$. It is invariant under the gauge transformation discussed earlier.
The $\mathbb{Z}_2$ gauge fields for the two layers, $u_{1,ij}$ and $u_{2,ij}$, are decoupled, leading to a $\mathbb{Z}_2\times\mathbb{Z}_2$ topological order described by 4-component Abelian Chern-Simons theory\cite{Wen1992} characterized by matrix ${\bf K}=\bpm0&2\\2&0\epm\oplus\bpm0&2\\2&0\epm$. However, the Goldstone mode of the N\'eel order
\bea\label{goldstone mode}
n^x+i n^y\sim b_{\vec k=(\pi,\pi)}\sim\sum_i(-1)^{i_x+i_y}f^\dagger_{2 i}f_{1i}
\eea
is not a gauge-invariant quantity, but instead an anyon obeying mutual semion statistics with the vison in each layer. More precisely, the above Goldstone mode carries the gauge charge for the $\mathbb{Z}_2$ gauge field from each layer. Incorporating the gapless anyon $b$ in (\ref{goldstone mode}) into the low energy description, the effective field theory for this algebraic spin liquid reads
\bea\notag
&\mathcal{L}_\text{ASL}=\sum_{I,J}\frac{\epsilon^{\mu\nu\rho}}{4\pi}a_\mu^I{\bf K}_{I,J}\partial_\nu a_\rho^J-\sum_{\alpha,I}\frac{\epsilon^{\mu\nu\rho}}{2\pi}A^\alpha_\mu{\bf q}^\alpha_I\partial_\nu a_\rho^I\\
&+
|(-\imth\partial_\mu-2A^s_\mu-a^1_\mu-a^2_\mu+a^3_\mu-a^4_\mu)^2b|^2+\cdots
\label{field theory}
\eea
where $A_\mu^{\alpha=c,s}$ label the charge and pseudo-spin external gauge fields, and \bea
{\bf q}_c=(2,0,2,0)^T,~~~{\bf q}_s=(2,0,-2,0)^T.
\eea
are the charge and pseudo-spin vectors\cite{Wen1992} for the Chern-Simons theory. 

%Hence this state features a $\mathbb{Z}_2$ gauge theory for the $f_{1i}$ fermions, which is decoupled from another $\mathbb{Z}_2$ gauge theory for the $f_{2i}$ fermions. The $U(1)_c$ and $U(1)_s$ symmetries (eq 4) of these fermions are preserved while the layer exchange symmetry $\hat{X}$ (eq 5) is spontaneously broken. Given these observations, we claim that we have a $\mathbb{Z}_2\times\mathbb{Z}_2$ topological order in this case. 

\noindent
{\it (ii)} The N\'eel vector lies in-plane, e.g. ${\bf n}\perp\hat z$ with
\begin{align}
   n^+\equiv n_i^x+in_i^y =(-1)^{r_{ix}+r_{iy}}\expval{f^\dagger_{1 i}f_{2 i}}\neq0
\end{align}
Unlike $n^z$, the in-plane components, $n^x$ and $n^y$ are not gauge invariant as the local gauge transformation maps $n^{x(y)} \rightarrow -n^{x(y)}$. However, a non-local gauge invariant correlator can be defined \cite{Nica_arXiv2022}.
\begin{eqnarray}
\mathcal{C}^{x(y)}(r,r') = \langle n^{x(y)}(r) B(r,r') n^{x(y)}(r') \rangle,
\label{Eq:Hbbr_crlt}
\end{eqnarray}
where the gauge string for fermions, $B(r,r')=  \prod_{ij \in (r,r')} u_{1 i j}u_{2 i j}$, connects operators at the end sites $(r,r')$. The value of $\mathcal{C}^{x(y)}(r,r')$ is the same in all gauge choices. Therefore, the ground state, symmetrized over all gauge configurations through the projection procedure, also has the same value of $\mathcal{C}^{x(y)}(r,r')$, signifying a string order parameter. Physically, the long-range string order corresponds to the condensation of anyon $b$ in the field theory (\ref{field theory}), hence breaking the gauge group down to $\mathbb{Z}_2$ via the Higgs mechanism\cite{Hansson_AnnPhys2004}. 

An alternative way to understand the gauge structure is to notice the following local order parameter for the in-plane N\'eel order
\bea
S^+_{\expval{i,j}}\equiv\expval{n_i^+B(i,j)n_j^+}
\eea
for a pair of nearest neighbor sites $\expval{i,j}$. Due to the mutual braiding phase of $e^{i\pi}$ between a vison and a fermion in each layer, a vison from layer 1 (or 2) is nothing but a vortex for the above local order parameter, since $S^+_{\expval{i,j}}$ acquires a $e^{\pm2\pi i}$ phase as it travels around a vison from layer 1 (2). The logarithmic confinement of vortices in the in-plane Neel phase suggest that the vison from layer 1 (or 2) is confined, therefore reducing the $Z_2\times Z_2$ gauge group down to $Z_2$. A similar conclusion can be drawn if the N\'eel vector has both in-plane and $\hat z$ components. 

%Here, the gauge degrees of freedom are not decoupled, as can be seen by the order parameter written in terms of the Majoranas: $\frac1V\sum_i(-1)^i\expval{c_{1i}^xc_{2i}^x+c_{1i}^yc_{2i}^y})$, it can be seen that this is not gauge invariant under (eq ?). Upon gauge transformation at site i on layer 1, another gauge transformation at the same site is needed on layer 2, indicating the locking of the gauge degrees of freedom in the two layers. This locking of gauges suggests a $\mathbb{Z}_2$ topological order.
%Correlators for the in-plane N\'eel vector components can be made gauge invariant by inserting a gauge string $\rho(r,r')=\prod_{\expval{ij} \in (r,r')}u_{1,ij}*u_{2,ij}$, leading to a gauge invariant correlator:$\langle N^x_r\rho(r,r') N^x_{r'}\rangle= \langle (c_{1r}^xc_{2r}^x+c_{1r}^yc_{2r}^y)\rho(r,r')(c_{1r'}^xc_{2r'}^x+c_{1r'}^yc_{2r'}^y)$. Now we have physical, although non-local correlators that are defined for our model, further strengthening the topological nature of the ground state. 

In general, the ground state can have both non-zero out-of-plane ($n^z$) and in-plane ($n^x, n^y)$ components. The term, `magnetic fragmentation' is coined for phases that display a coexistence of a local Landau-type order parameter and a non-local topological order \cite{Brooks_PRX2014, Petit_NatPhys2016, Lefrancois_NatComm2017, Zorko_PRB2019, Mauws_2018,Wang2022}. Magnetic fragmentation is theoretically predicted\cite{Brooks_PRX2014} and experimentally observed\cite{Petit_NatPhys2016} in spin ice materials such as Nd$_2$Zr$_2$O$_7$ where a local AFM order coexists with a spin liquid with FM correlations. 
%To best of our knowledge, our model provides the first solvable model of magnetic fragmentation. %Moreover, a symmetry breaking field along the $z$-direction can be included to eliminate the in-plane component of the order parameter. This changes the topological order from $\mathbb{Z}_2$ to $\mathbb{Z}_2 \times \mathbb{Z}_2$.   
The conclusions we draw from the Hubbard model rely on mean-field order parameters. Next, we underpin these results by a perturbative analysis.

\noindent 
\textbf{Perturbative analysis in the limit of large inter-layer exchange.} We corroborate the results of the Hubbard model (Eq.~\ref{eq:Hcf}) by considering the bilayer in the large-$J$ limit, without reference to the Majorana representation (Eq.~\ref{eq:H}), on a torus. We introduce effective pseudo-spin and orbital DOF appropriate to this limit. We next derive effective models on the large-$J$ GS manifold, to fourth order in the intra-layer coupling $K$. By analogy to the Hubbard model, we distinguish between cases with a) $\mathbb{Z}_{2}$ and b) $\mathbb{Z}_{2} \times \mathbb{Z}_{2}$ topological order. For a), we show that the GS manifold is a state of uniform $\pi$ flux, with a finite $\mathcal{C}^{x(y)}$ correlator (Eq.~\ref{Eq:Hbbr_crlt}). We also demonstrate that the GS manifold has four-fold topological degeneracy and that the visons are confined. For b), we also obtain a GS with uniform $\pi$ flux, which has sixteen-fold topological degeneracy and deconfined vison excitations. These results naturally lead to the conclusion that the two phases are separated by a topological phase transition.    
\\

\noindent \textit{Effective degrees of freedom.} We first introduce the effective pseudo-spin and orbital DOF. For $K=0$ and finite FM inter-layer interactions ($J<0$), the spins on overlapping sites form GS doublets $\ket{\uparrow_1\uparrow_2},\ket{\downarrow_1\downarrow_2}$. These can be represented by a bilayer pseudo-spin 
  
\begin{eqnarray}
\eta^z_{i}&=&\frac{1}{4}(\sigma_{1i}^z+\sigma_{2i}^z) \nonumber \\
\eta^\pm_{i}&=& \frac{1}{4} \sigma_{1i}^\pm\sigma_{2i}^\pm, 
\end{eqnarray}
%%%
\noindent obeying an SU(2) algebra. In addition, the orbital DOF for each pair of overlapping sites form a four dimensional Hilbert space, corresponding to one singlet and three triplet configurations. To represent these states, we introduce the inter-layer orbital operators ${q}_i^\gamma = \tau_{1i}^{\gamma}\tau_{2i}^{\gamma}, ~\gamma=x,y,z$, which mutually commute as $[{q}_i^\alpha, {q}_i^\beta]=0$. The four orbital states can be labeled by the three eigenvalues $q^{\gamma}_{i}=\pm 1$, constrained to obey $\prod _{\gamma} q_i^{\gamma} = -1$. The Hilbert space thus includes all states of the form

\noindent \begin{align}
\ket{ \{\eta^{z} \},  \{ q^{\gamma} \}} = \ket{\{\eta^{z}_{i}\}} \otimes \ket{\{( q^{x}_{i}, q^{y}_{i}, q^{z}_{i} )\}},
\end{align}

\noindent with the implicit local constraint. Pairs of nearest-neighbor $q^{\gamma}_{i/j}$ define the bond variables

\noindent \begin{align}
\rho^{\gamma}_{ij} = q^{\gamma}_{i}q^{\gamma}_{j}
\end{align}

\noindent which take on values of $\pm 1$ for $\gamma \in \{x,y,z\}$, while they are trivially equal to 1 for additional identity bonds, labeled by $\rho^{\gamma=\rm{I}}\equiv 1$. To any $\{q^{\gamma}_{i} \}$ configuration, we can associate a unique $\{ \rho^{\gamma}_{ij}\}$ bond configuration, while the converse is not true.  Orbital states like $\ket{\phi}_{0} = \ket{\forall ~q^{\gamma}_{i} = -1}$, which have uniform $\rho^{\gamma}_{ij}=1$, play an important role in all subsequent discussions. 

\begin{figure}[t]
\includegraphics[width=7.9cm]{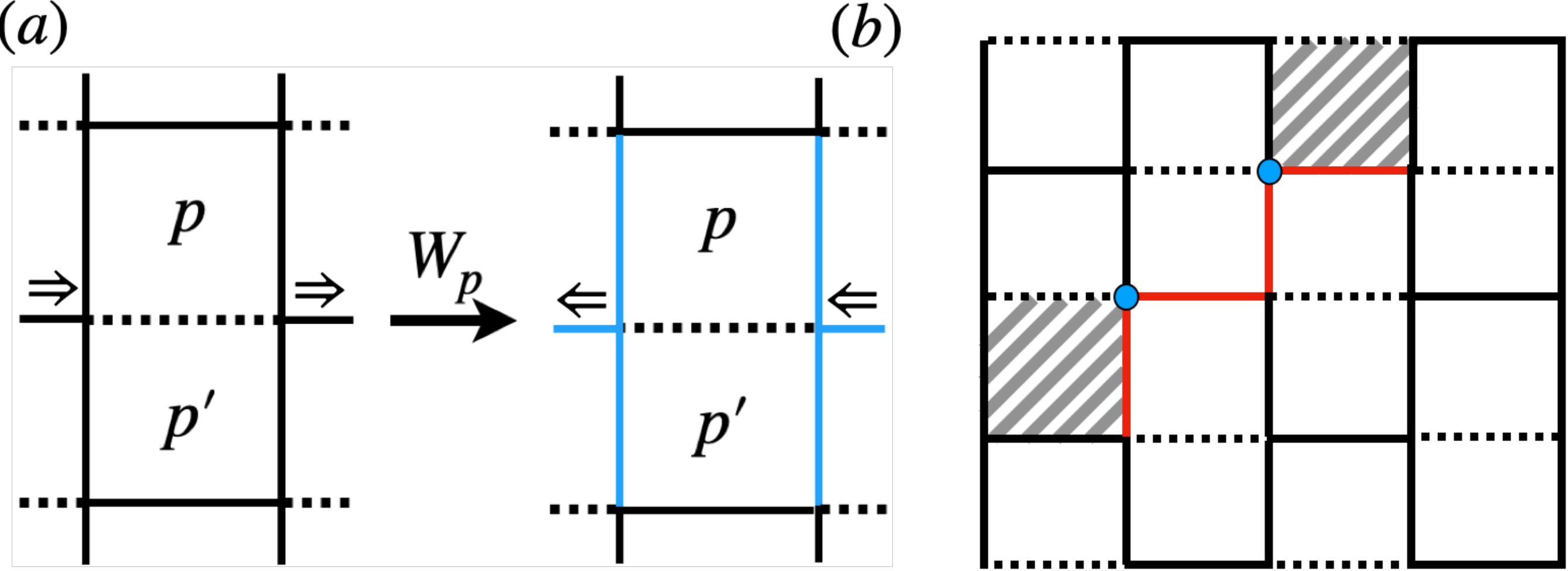}
\caption{Illustration of the pseudo-spin and bond ($\eta-\rho$) configurations. (a) Effect of flux operator $W_{\nu p}$ on a state with fixed $\rho^{\gamma}_{ij}$ bonds. $W_{\nu p}$ changes the signs of the bonds marked in blue. It also changes the signs of $\eta^{x/y}$ pseudo-spins on the identity (dashed) bond. (b) String defect, with red lines indicating bonds with signs opposite to the background bonds, marked in black. The string shown here is created by operating with $\tau^{x}_{\nu i}$ on any orbital state along the sites marked with blue dots. When operating on the GS in Eq.~\ref{Eq:Hg4_GS}, the $\tau$'s change the fluxes on the hashed plaquettes, since they anti-commute with $W_{\nu p/p'}$. Consequently, this open string terminates with a pair of visons.}
\label{Fig:2}
\end{figure}

Defects in $\ket{\phi}_{0}$ take the form of strings of negative bonds as shown in Fig.~\ref{Fig:2}~(b). Defects in both pseudo-spin and $\rho^{\gamma}_{ij}$ bonds are introduced by operating with the flux operators $W_{\nu p/p'}$. Each of these flips the pseudo-spin components along $x/y$ as $\eta^{x/y}_{n,k} \rightarrow - \eta^{x/y}_{n,k}$ in  the corresponding unit cell (\ref{Fig:2}~(a)). Each also changes the signs of all six $\rho^{\gamma}$ bonds connected with sites $n,k$. Note that any string defect cannot be eliminated by application of $W_{\nu p/p'}$ operators.
\\

\noindent \textit{Effective Hamiltonian.} The effective Hamiltonian $H_{\eta-\rho}$, projected onto the $K=0$ GS manifold, reads

\noindent \begin{align}
H_{\eta-\rho} = H_{g_{2}} + H_{g_{4}},
\end{align}

\noindent  where

\noindent 
\begin{widetext}
\begin{align}
    H_{g_2}= & g_2  \bigg[ \sum_{\langle ij \rangle}\eta_i^z\eta_j^z+ \sum_{\langle ij \rangle_{\gamma}} \frac{1}{2}(\eta^+_i\eta^-_j+\eta^-_i\eta^+_j)\rho^\gamma_{ij} \bigg] + \sum_{i} (-1)^{i_{x}+i_{y}} \left( h_{x} \eta^{x}_{i} + h_{z} \eta^{z}_{i} \right),
    \label{Eq:Hg4}
\end{align}
\end{widetext}

\noindent is obtained at second order in $K$  ($g_2=K^2/4J$). Note the distinction between NN, in-plane pseudo-spins connected via variable and trivial identity $\rho^{\gamma}_{ij}$ bonds, respectively. We introduce small perturbations $h_{x/z} > 0$ to explicitly break the continuous symmetry, enforcing the staggered pseudo-spin configurations along $x/z$, respectively. As it turns out, these respectively correspond to $\mathbb{Z}_{2}$ and $\mathbb{Z}_{2} \times \mathbb{Z}_{2}$ topological order. The fourth-order contribution is

\noindent \begin{align}
    H_{g_4}=g_4\left[\sum_{\nu, p} \eta^\square_{p} W_{\nu p}+\sum_{\nu, p'} \eta^\square_{p'} W_{\nu p^\prime} \right],
\end{align}

\noindent where $g_4=K^4/J^3$. $\eta^\square_{p}$ and $\eta^\square_{p'}$ include linear combinations of products of $\eta$ and $q^{\gamma}$ operators around $p/p'$ plaquettes. $H_{g_{2}}$ commutes with all flux operators for $h_{x} \rightarrow 0$, while this always holds for $H_{g_{4}}$.
Technical details and derivations related to the effective Hamiltonian and the following sections are relegated to SM. 
\\

\noindent \textit{$\mathbb{Z}_{2}$ topological order}. We consider $H_{\eta-\rho}$ with $h_{x}>0$ and $h_{z}=0$, on a torus. Exact diagonalization calculations indicate that the GS manifold of $H_{g_{2}}$ includes $\ket{\eta_{xx},\phi_{0}}$, with $\eta_{xx}$ denoting a finite staggered pseudo-spin along $x$ (see SM). Importantly, any configuration with string defects in the $\rho^{\gamma}$ bonds are gapped, with an energy cost which scales as the string length.%, as shown in SM. 

We next consider the evolution of the GS manifold at $H_{g_{4}}$ level for $h_{x} \ll g_{4} \ll g_{2}$. In SM, we show that exact diagonalization calculations indicate $H_{g_{4}}$ projects $\ket{\eta_{xx},\phi_{0}}$ onto a state of uniform $\pi$ flux per plaquette:

%\noindent \begin{widetext}
\noindent \begin{align}
  \ket{\Psi_{\rm{GS}}} = & \prod_{\nu, p, p'} \frac{\left(1-W_{\nu p}  \right) \left(1-W_{\nu p'}  \right)}{4} \ket{\tilde{\eta}_{xx}; \phi_{0}}  + O\left(\frac{h_{x}}{g_{4}} \right). 
  \label{Eq:Hg4_GS}
\end{align}
%\end{widetext} 
Note that $\tilde{\eta}_{xx}$ is a state of staggered pseudo-spins including  corrections at both $H_{g_{2}}$ and $H_{g_{4}}$ levels. This is a state of definite $\pi$ flux since $W_{\nu p} (1 - W_{\nu p}) = - (1 - W_{\nu p})$. Moreover, all $W_{\nu p/p'}$ commute with the operator 

\noindent \begin{align}
\mathcal{D}^{x}_{ij}= & \eta^{x}_{i} \left( \prod_{i'j' \in C_{ij}} \rho^{\gamma}_{ij} \right) \eta^{x}_{j}.
\end{align}

\noindent Consequently, the latter has a finite expectation value in $\ket{\Psi_{\rm{GS}}}$ for any pair of $i,j$, reflecting a locking of pseudo-spin and $\rho^{\gamma}_{ij}$ bond configuration. %In SM, we show that 
Moreover, $\mathcal{D}^{x}_{ij}$ is equivalent to a gauge-invariant correlator of the Hubbard model corresponding to an in-plane N\'eel vector~(Eq.~\ref{Eq:Hbbr_crlt}). 

Note that any state with an open string defect, obtained by first including strings in $\phi_{0}$, involves one or more visons on the plaquettes at each end~(\ref{Fig:2}~(b)). As the energy of this excitation depends on the string length and diverges for infinite vison separation, it follows that the latter are confined in an infinite system.  
\\

\noindent \textit{$\mathbb{Z}_{2} \times \mathbb{Z}_{2}$ topological order.} In this case, we consider $H_{\eta-\rho}$ with $h_{z} > 0$ and $h_{x}=0$. By analogy with the case with $\mathbb{Z}_{2}$ topological order, the GS manifold of $H_{g_{2}}$ now includes $\ket{\eta_{zz}, \phi_{0}}$, where $\eta_{zz}$ indicates finite staggered pseudo-spins along $z$. Similarly, any open string defects are gapped. However, in contrast to the case for $\mathbb{Z}_{2}$ topological order, the gap for these excitations remains finite for arbitrary string length, in the infinite-system size. In the same limit, states with strings forming non-contractible loops become degenerate with $\ket{\phi_{0}}$  for $h_{z} \gg g_{2/4}$. 

The effect of $H_{g_{4}}$ is analogous to the case with $\mathbb{Z}_{2}$ topological order. Consequently, the GS has the form in Eq.~\ref{Eq:Hg4_GS}, with the replacement $\ket{\tilde{\eta}_{xx}; \phi_{0}} \rightarrow \ket{\tilde{\eta}_{zz}; \phi_{0}}$, indicating a surviving pseudo-spin staggering. While the two-point correlator for the $\eta^{z}$ pseudo-spins is always finite, $\mathcal{D}^{x}_{ij}$ vanishes for infinite separation as in the Hubbard model with $\mathbb{Z}_{2} \times \mathbb{Z}_{2}$ topological order. 

Since the energy cost of an open string remains finite even in an infinite-size system, the visons are deconfined. Similarly, as states with non-contractible loops of negative $\rho^{\gamma}$ bonds become degenerate with the GS, in the limit of infinite system-size, the latter acquires additional topological degeneracy. Consequently, this entails a sixteen-fold topological degeneracy on the torus, consistent with $\mathbb{Z}_{2} \times \mathbb{Z}_{2}$ topological order.
\\
 
\noindent
{\bf Conclusion and outlook.} We have studied a bilayer adaptation of a QSL model on a square lattice with Kitaev-type interactions. We have shown that the low energy model exhibits an AFM Mott transition which corresponds to magnetic fragmentation in terms of the original DOF. We have corroborated these results by a perturbative calculation for the topological degeneracy, which is consistent with field theory analysis. The analysis we have presented here may be of particular value as a largely tractable yet highly non-trivial instance of magnetic fragmentation.

Interesting future directions include examining the role of fluctuations on the emergent symmetry which may reduce the ground state manifold via order by disorder\cite{Shender_obdo,Henley_obdo,Green_ARCMP2018}. Another direction is to generalize our mechanism for fractionalized Goldstone modes in bilayer systems to multilayer systems with larger emergent symmetries, such as $SU(N)$. 

The study of moir\'e superlattices of QSLs is another intriguing direction\cite{Luo_PRB2022,Nica_arXiv2022}. Our work suggests that manifold new phenomena arising from a combination of emergent symmetry and strong interactions are awaiting discovery here, providing a new vista on strongly correlated magnetism. 

%We studied a bilayer adaptation of a QSL model on a square lattice with Kitaev-type interactions. We showed that the low energy description is given by an SU(2) $\pi$-flux Hubbard model and the AFM Mott transition in the model correspond to a magnetic fragmentation transition in terms of spin and orbital degrees of freedom. We benchmarked these results by perturbative analysis and showed that topological degeneracy and magnetic fragmentation without the fractionalization of the spin. Interesting future directions include moir\'e superlattices of QSLs.

\noindent
{\bf Acknowledgements.} We thank Nandini Trivedi and Natalia Perkins for fruitful discussions. OE acknowledge support from NSF Award 
No. DMR 2234352. YML is supported by NSF under Award No. DMR 2011876. EMN acknowledges support by NSF under Grant No. DMR-2220603. This work in part supported by the Deutsche Forschungsgemeinschaft (DFG) via SFB 1143 (project-id 247310070) and cluster of excellence ct.qmat (EXC 2147, project-id 390858490).


%\bibliography{references.bib}
\end{document}

\title{Supplementary material for ``Emergent symmetry and fractionalized Goldstone modes in a bilayer quantum spin liquid"}
\author{Aayush Vijayvargia$^1$, Emilian Marius Nica$^{1,2}$, Roderich Moessner$^3$, Yuan-Ming Lu$^{4}$, Onur Erten$^1$}
\affiliation{$^1$Department of Physics, Arizona State University, Tempe, AZ 85287, USA \\ $^{2}$Department of Physics and Astronomy, Rice University, 6100 Main St, Houston 77005 TX, USA \\ $^3$Max-Planck-Institut f\"ur Physik komplexer Systeme, N\"othnitzer Strasse 38, 01187 Dresden, Germany\\ $^4$ Department of Physics, The Ohio State University, Columbus OH 43210, USA}

%\author{Aayush Vijayvargia$^1$, Emilian Marius Nica$^1$, Roderich Moessner$^2$, Yuan-Ming Lu$^{3}$, Onur Erten$^1$}
%\affiliation{$^1$Department of Physics, Arizona State University, Tempe, AZ 85287, USA \\ $^2$Max-Planck-Institut f\"ur Physik komplexer Systeme, N\"othnitzer Strasse 38, 01187 Dresden, Germany\\ $^3$ Department of Physics, The Ohio State University, Columbus OH 43210, USA}
\maketitle

\section{Majorana fermion representation}

The Hamiltonian of the bilayer is  
\begin{align}
  H=H_K+H_J
  %=-\sum_{\langle ij \rangle_\gamma,\nu }K_\nu(\sigma^x_{\nu i}\sigma^x_{\nu j} +\sigma^y_{\nu i}\sigma^y_{\nu j})(\tau_{\nu i}^{\gamma}\tau_{\nu j}^{\gamma})+J\sum_i\sigma^{z}_{1i}\sigma^{z}_{2i}
    \label{eq:H}
\end{align}
where $H_K$ and $H_J$ are the intra- and inter-layer terms, respectively. $H_K$ has the form  \cite{Nakai_PRB2012},
%The single-layer model we have considered here is a Kitaev like spin-orbital model \cite{Nakai_PRB2012}. We have four-dimensional gamma matrices obeying the Clifford algebra on each site of a square lattice. The Hamiltonian is: 
\begin{equation}
    H_K=-K\sum_{\langle ij \rangle_{\gamma},\nu}(\Gamma_{\nu i}^{\gamma}\Gamma_{\nu j}^{\gamma}+\Gamma_{\nu i}^{\gamma5}\Gamma_{\nu j}^{\gamma5})
    \label{eq:HK}
\end{equation}
where $\Gamma^{\gamma}$ matrices obey the Clifford algebra, $\{\Gamma^\alpha,\Gamma^\beta \} =\delta_{\alpha,\beta}$ and $\Gamma^{\alpha \beta}=[\Gamma^\alpha,\Gamma^\beta]$.
Using 
%We use two sets of Pauli matrices to re-express the above Hamiltonian: 
$\Gamma^{\alpha}=-\sigma^y\otimes \tau^{\alpha} \quad (\alpha=x,y,z)$, $\Gamma^4=\sigma^x\otimes\mathbb{I}_2$ and $\Gamma^5=-\sigma^z\otimes\mathbb{I}_2$, $H_K$ can be re-cast as

\begin{eqnarray}
H_K=-\sum_{\langle ij \rangle_\gamma,\nu }K_\nu(\sigma^x_{\nu i}\sigma^x_{\nu j} +\sigma^y_{\nu i}\sigma^y_{\nu j})(\tau_{\nu i}^{\gamma}\tau_{\nu j}^{\gamma})
\end{eqnarray}
where $\sigma (\tau)$ Pauli matrices act on spin(orbital) degrees of freedom (DOF). Here $\tau^{\gamma} = \tau^x,\tau^y,\tau^z,\mathbb{I}$ and $\gamma=1,2,3,4$ correspond to the 4 inequivalent bonds at each site. We introduce a Majorana fermion representation via

\begin{align}
    \Gamma^{\alpha}_j=ib_j^{\alpha}c_j, \qquad \Gamma^{\alpha \beta}_i=ib_i^{\alpha}b^{\beta}_i
    \label{eq. majorep}
\end{align} 
and relabel $b_i^5\rightarrow c_i^x$ and $c_i\rightarrow c_i^y$ for convenience to obtain
\begin{equation}
    H_K=K\sum_{\langle ij \rangle_{\gamma}}\text{i} u_{\nu,ij}[c^x_{\nu i}c^x_{\nu j}+c^y_{\nu i}c^y_{\nu j}]
\end{equation}
where $u_{\nu,ij}=ib^{\gamma}_{\nu i}b^{\gamma}_{\nu j}$. 
The inter-layer interaction Hamiltonian, $H_J=J\sum_i\Gamma^5_{1i}\Gamma^5_{2i}$ can be similarly expressed as
\begin{eqnarray}
    H_J & = & J\sum_i\sigma^{z}_{1i}\sigma^{z}_{2i}
    %&=&-J^z\sum_i\Gamma^5_{1i}\Gamma^5_{2i} 
    \nonumber \\
    &=&-J\sum_ic^{x}_{1,i}c^{y}_{1,i}c^{x}_{2,i}c^{y}_{2,i}\label{eq:inter layer}
\end{eqnarray}
Note that the fermion Hilbert space per site is $2^3=8$-dimensional, twice the size of the 4-dimensional physical Hilbert space expanded by Pauli matrices $\vec\sigma$ and $\vec\tau$. To faithfully represent the physical system with the Majorana fermion degrees of freedom, we have to enforce the following onsite constraint
\bea
D_{\nu j}=\imth b^1_{\nu j}b^2_{\nu j}b^3_{\nu j}b^4_{\nu j}c^x_{\nu j}c^y_{\nu j}=1,~~~\forall~j,\nu.\label{onsite constraint}
\eea

\section{Solution of the intra-layer Hamiltonian}
According to Lieb's theorem\cite{Lieb_PRL1994}, the ground state of each layer of $H_K$ lies in the $\pi$-flux sector. Such states  can be obtained by choosing gauge configurations where every x-bond (see Fig. 1(a) in the main text) has $u_{ij}=-1$, while all remaining bonds have $u_{ij}=1$. 
\begin{suppfigure}[t]
\includegraphics[width=8.0 cm]{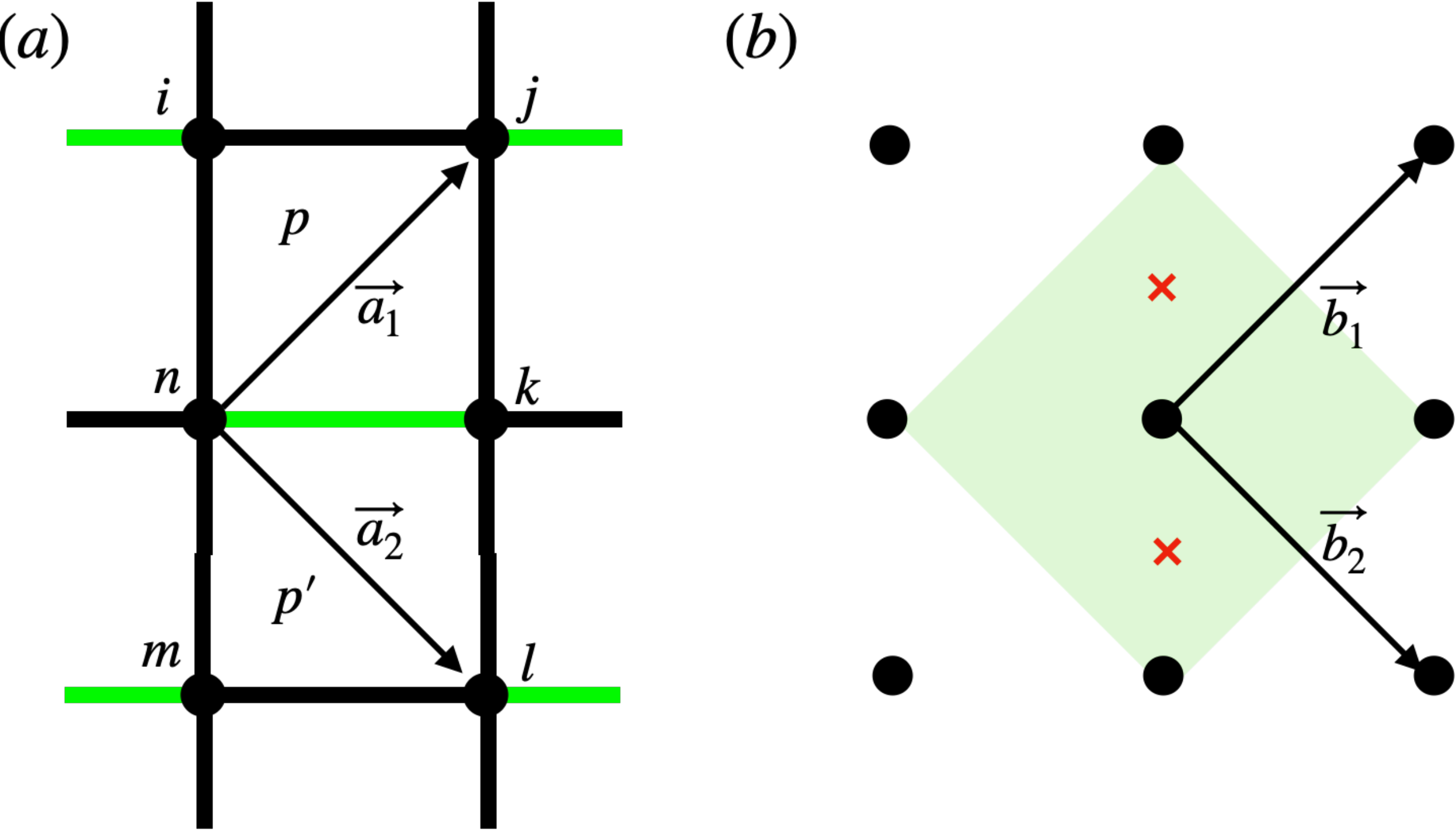}
\caption{(a) Lattice vectors $\vec{a}_{1,2}=(1,\pm1)$. The unit cell includes two inequivalent plaquettes labeled $p$ and $p'$. Note the specific gauge choice adopted here, where green bonds and black bonds correspond to $u_{ij}=-1$ and $u_{ij}=1$, respectively. (b) Brillouin zone (BZ) (green square) with reciprocal vectors $\vec{b}_{1,2}=\pi(1,\pm1)$. Red crosses indicate the two inequivalent Dirac points at $(0,\pm\frac{\pi}{2})$.}
\label{Fig:S1}
\end{suppfigure}
The unit cell includes two inequivalent square plaquettes with primitive lattice vectors $a_{1,2}=(1,\pm 1)$ and corresponding reciprocal vectors $b_{1,2}=\pi(1,\pm 1)$ (in units of $a=1$) as shown in Supplementary figure \ref{Fig:S1}. Performing a Fourier transform  $f_{(A,B)i}=\Sigma_k e^{-ikr_i} f_{(A,B)k}$, we obtain
\begin{align}
  H_K= \sum_{k \in 1^{\rm st} ~{\rm BZ}} \begin{pmatrix}
    f_{Ak}^{\dag} && f_{Bk}^{\dag}
    \end{pmatrix}
    \begin{pmatrix}
    0 && 4K(\cos k_y +i\sin k_x) \\
    4K(\cos k_y -i\sin k_x) && 0
    \end{pmatrix}
    \begin{pmatrix}
    f_{Ak}\\
    f_{Bk}
\end{pmatrix}
\label{eq:Hkk}
\end{align}
%Diagonalizing eq.~\ref{eq:Hkk}, we obtain 
with a dispersion $E_K=\pm 4K\sqrt{\sin^2{k_x}+\cos^2{k_y}}$. $E_K$ has two inequivalent Dirac points at $(0,\pm\frac{\pi}{2})$ as shown in Supplementary figure \ref{Fig:S1}.

\section{Symmetry and topological properties}

\subsection{Symmetry of the full Hamiltonian}

We first consider the symmetry of the single layer Hamiltonian (\ref{eq:HK}). 
%Below we write down the symmetry generators in the Majorana representation. 
%First we notice a 
The (spinless) time reversal symmetry operator $\bst=\mathcal{K}$ is simply implemented by the complex conjugation $\mathcal{K}$. In addition, we have a unitary $U(1)_\nu\rtimes \mathbb{Z}_2^\bsc$ symmetry, generated by $U(1)$ ``charge'' rotations in each layer $\nu$
\bea
U(1)_\nu=\{e^{\imth\theta\bsq_\nu}|\bsq_\nu=\sum_j\frac{\imth c_{\nu j}^xc^y_{\nu j}+1}2,~0\leq\theta<2\pi\},
\eea
and the particle-hole symmetry in layer $\nu$
\bea
\bsc_\nu=\prod_j(\imth c^y_{\nu j}b^\gamma_{\nu j}),\label{phs:each layer}
\eea
where we can choose an arbitrary (but fixed) bond direction $1\leq\gamma\leq4$. Note that all these symmetries commute with the onsite constraint (\ref{onsite constraint}).

Next we take the inter layer interaction term (\ref{eq:inter layer}) into account to analyze the symmetry of the full Hamiltonian (\ref{eq:H}). Note that the particle-hole symmetry (\ref{phs:each layer}) in each layer does not commute with the onsite constraint (\ref{onsite constraint}), and hence ceases to be a symmetry of the full Hamiltonian (\ref{eq:H}). Instead, the following particle-hole symmetry for both layers
\bea
\bsc\equiv\bsc_1\bsc_2=\prod_{j,\nu}(\imth c^y_{\nu j}b^\gamma_{\nu j})
\eea
remain as a symmetry of the full Hamiltonian. Moreover, there is a layer exchange symmetry generated by
\bea
\mathcal{X}=\prod_i\big[\frac{1+c^x_{1,i}c^x_{2,i}}{\sqrt2}\cdot\frac{1+c^y_{1,i}c^y_{2,i}}{\sqrt2}\cdot\prod_\gamma \frac{1+b^\gamma_{1,i}b^\gamma_{2,i}}{\sqrt2}\big]
\eea

As a result, the symmetry of the full Hamiltonian is given by 
\bea
G=\big[U(1)\times U(1)\times \mathbb{Z}_2^\bst\big]\rtimes\big[\mathbb{Z}_2^{\mathcal{X}}\times \mathbb{Z}_2^\bsc\big]
\eea

The symmetry of the full Hamiltonian can be conveniently understood in the basis of complex fermions $f_{\nu,j}$ defined as
\bea\label{complex basis}
f_{\nu,j}\equiv\frac{c^x_{\nu,j}-\imth c^y_{\nu,j}}2\Longrightarrow\{f_{\nu,i},f_{\mu,j}^\dagger\}=\delta_{i,j}\delta_{\mu,\nu}
\eea
In this basis, the $U(1)$ symmetries for each layer can be recombined into ``charge'' $U(1)_c$ and pseudo-spin (i.e. layer) $U(1)_s$ rotations:
\bea
U(1)_c:~~~f_{\nu,j}\rightarrow e^{\imth\theta}f_{\nu,j};~~~\bsc f_{\nu,j}\bsc^{-1}=f^\dagger_{\nu,j};\\
U(1)_s:~~~f_{\nu,j}\rightarrow e^{\imth\kappa\theta}f_{\nu,j};~~~\mathcal{X}f_{\nu,j}\mathcal{X}^{-1}=\kappa f_{\nu',j},
\eea
where $\kappa=-1,1$ for $\nu=1,2$ respectively and $\nu'=\nu +1 \rm (mod 2)$.
If we represent the pseudo-spin-1/2 complex fermions using the following matrix
\bea\label{complex basis:matrix}
F_j\equiv\bpm f_{1,j}&f_{2,j}^\dagger\\f_{2,j}&-f^\dagger_{1,j}\epm
\eea
the pseudo-spin and charge rotations correspond to left and right rotations on the matrix $F_j$:
\bea
&R(U_\theta^s)=e^{\imth\theta Z_L};~~~R(\mathcal{X})=\imth Y_L;\\
&R(U_\theta^c)=e^{\imth\theta Z_R};~~~R(\bsc)=X_R.
\eea
where $\vec\sigma_{L/R}\equiv(X_{L/R},Y_{L/R},Z_{L/R})$ stands for Pauli matrices multiplied to the left/right of matrix $F_j$ in (\ref{complex basis:matrix}). Now it is clear that the full symmetry group can be written as
\bea\label{full symmetry}
G=O(2)_c\times O(2)_s\times \mathbb{Z}_2^\bst
\eea
where $O(2)=U(1)\rtimes \mathbb{Z}_2$ is the rotation group for either the charge or pseudo-spin degrees of freedom.

\subsection{Emergent symmetry for a fixed flux configuration}

One type of gapped       excitations in model (\ref{eq:H}) are the visons\cite{Kitaev_AnnPhys2006}, i.e. $\pi$ flux excitations in each plaquette. More precisely, Majorana fermions $\{c^{x,y}_{\nu,i}\}$ hop in a background flux, where the flux in the plaquette $\expval{i,j,k,l}$ is given by
\bea\label{flux operator}
F_{\nu,ijkl}=u_{\nu,ij}u_{\nu,jk}u_{\nu,kl}u_{\nu,li}=\pm1
\eea
In a fixed flux configuration where both layers share the same flux in every plaquette, we can always choose a gauge so that $u_{\nu,ij}\equiv u_{ij}=\pm1$ are fixed as constants. Therefore the physical spectrum with fixed fluxes (i.e. without vison excitations) is given by a Hubbard model of Majorana fermions $\{c^{x,y}_{\nu,i}\}$ with a hopping strength $K$ and a Hubbard interaction strength $J$. The effective symmetry for a fixed flux configuration is generated by the spinless time reversal symmetry
\begin{equation}\label{time reversal}
\bst=\prod_{j,\nu}(\imth c^x_{\nu j}c^y_{\nu j})^{j_x+j_y}\cdot\mathcal{K}    
\end{equation}
and the well-known unitary $SO(4)=SU(2)\times SU(2)/\mathbb{Z}_2$ symmetry of a 2-flavor Hubbard model\cite{Yang1990}, where the layer index $\nu$ plays the role of the psuedo-spin. We have used $j=(j_x,j_y)\in\mbz^2$ to label a site on the square lattice. Therefore the emergent symmetry group for a fixed flux configuration is given by
\bea
G^\prime=SO(4)\times \mathbb{Z}_2^\bst
\eea
It is most convenient to represent the symmetry in the $u_{\nu,ij}\equiv u_{ij}=\pm1,~\forall~\nu=1,2$ gauge, where the $SO(4)$ symmetry is simply implemented by an $SO(4)$ unitary rotation on the Majorana basis
\bea\label{majorana basis}
\psi_j\equiv(c^x_{1,j},c^y_{1,j},c^x_{2,j},c^y_{2,j})^T
\eea
since the fixed-flux Hamiltonian in this gauge is written as
\bea\label{fixed flux ham}
\hat H_{\text{fixed flux}}=-K\sum_{\expval{i,j}}u_{ij}\sum_{\nu=1,2}\sum_{\alpha=x,y}\imth c^\alpha_{\nu,i}c^\alpha_{\nu,j}+J\sum_ic^x_{1,i}c^y_{1,i}c^x_{2,i}c^y_{2,i}
\eea
Meanwhile, under the spinless time reversal symmetry operation, the Majorana fermions (\ref{majorana basis}) transform as 
\bea
\psi_j\overset{\bst}\longrightarrow(-1)^j\psi_j
\eea
The $SO(4)$ symmetry can also be understood in the basis of complex fermions $f_{\nu,j}$ defined in (\ref{complex basis}), where $SU(2)_s$ pesudo-spin rotation and $SU(2)_c$ charge rotation are realized as left and right $SU(2)$ rotations on the matrix in (\ref{complex basis:matrix}):
\bea
F_j\equiv\bpm f_{1,j}&f_{2,j}^\dagger\\f_{2,j}&-f^\dagger_{1,j}\epm\rightarrow M_sF_jM_c^\dagger,~~~M_{c,s}\in SU(2)
\eea

Physically, at an energy scale much smaller than the vison gap, vison excitations are suppressed and the low energy subspace has a fixed flux configuration. As proven by Lieb\cite{Lieb1994}, a uniform $\pi$ flux per plaquette has the lowest energy and the low energy physics is described by Majoranas $\psi_j$ hopping in a uniform background of $\pi$ flux per plaquette. Therefore, the effective (or emergent) symmetry of the system below the vison excitation gap is exactly given by the symmetry of a fixed flux configuration as we described above. 

\subsection{Remnant symmetry of the N\'eel order}

In a Hubbard model on the square lattice, as the onsite repulsive Hubbard interaction strength $J/K$ increases beyond a critical value $J_c/K$, the system enters a N\'eel ordered phase with the long-range antiferromagnetic order. Now that we have obtained the symmetry of the full Hamiltonian (\ref{eq:H}) and of any fixed flux configuration, next we analyze the remnant (unbroken) symmetry of the N\'eel ordered ground state in our model (\ref{eq:H}). 

The N\'eel order in the Hubbard model is characterized by the staggered magnetization as its order parameter, represented by the N\'eel vector:
\bea\label{Neel vector}
\textbf{n} =\frac{1}{N}\sum_j{\bf n}_j,~~~{\bf n}_j = (-1)^{j_{x}+j_{y}}\langle f^\dagger_{\mu j}\boldsymbol \sigma_{\mu \nu}f_{\nu j}\rangle=(-1)^{j_x+j_y}\text{Tr}(F_j^\dagger{\boldsymbol \sigma}F_j)
\eea

Note that the N\'eel vector always changes sign when transformed under the spinless time reversal symmetry $\bst=\mathcal{K}$, therefore it always spontaneously breaks the antiunitary time reversal symmetry $\bst$. From (\ref{Neel vector}) it is clear that the N\'eel vector is invariant under charge rotations $M_c\in SU(2)_c$ in (\ref{complex basis}), therefore it always preserves the $O(2)_c$ subgroup of ``charge'' symmetries of the full Hamiltonian (\ref{eq:H}), or the $SU(2)_c$ subgroup of the emergent $G^\prime=SO(4)\times \mathbb{Z}_2^\bst$ symmetry for a fixed flux configuration. In fact, independent of the N\'eel vector orientation, in a fixed flux configuration, the N\'eel order always spontaneously breaks the emergent symmetry group $G^\prime=SO(4)\times \mathbb{Z}_2^\bst$ down to a subgroup
\bea
H^\prime=SU(2)_c\times U(1)_{\bf n}\rtimes \mathbb{Z}_2^{{\bst}_{\bf n}}
\eea
where in the complex fermion basis (\ref{complex basis:matrix}) we have defined
\bea\label{in-plane time reversal}
U(1)_{\bf n}=\{e^{\imth\theta\hat{\bf n}\cdot\vec\sigma_L}|0\leq\theta<\pi\},~~~R(\bst_{\bf n})\equiv(\imth\hat{\bf m}\cdot\vec\sigma_L)\cdot Y_LY_R\cdot\mathcal{K}
\eea
Here $\hat {\bf m}\parallel\hat z\times {\bf n}$ is an in-plane unit vector perpendicular to the unit N\'eel vector $\hat{\bf n}$.

Below we analyze three cases with different orientations of the N\'eel vector ${\bf n}$, where the symmetry group $G$ in (\ref{full symmetry}) of the full Hamiltonian is spontaneously broken down to different subgroups. 

(1) If the N\'eel vector ${\bf n}$ is parallel to the $\hat z$ axis, the $U(1)_s$ symmetry is preserved while the layer exchange $\mathcal{X}$ is broken. As a result, the remnant symmetry group is 
\bea
H({\bf n}\parallel\hat z)=O(2)_c\times U(1)_s\rtimes \mathbb{Z}_2^{\mathcal{X}\cdot\bst}
\eea
The associated order parameter manifold is given by
\bea
\mathcal{M}=G/H=\mbz_2
\eea

(2) If the N\'eel vector ${\bf n}$ lies in the plane perpendicular to the $\hat z$ axis, the $U(1)_s$ spin rotation is broken while a layer exchange symmetry $\mathcal{X}_{\bf n}$ defined by 
\bea
R(\mathcal{X}_n)=\imth\hat {\bf n}\cdot\vec\sigma_L
\eea
in the complex fermion basis (\ref{complex basis:matrix}) is preserved. As a result, the remnant symmetry group is 
\bea
H({\bf n}\perp\hat z)=O(2)_c\times \mathbb{Z}_2^{\mathcal{X}_{\bf n}}\times \mathbb{Z}_2^{\tilde\bst}
\eea
where the preserved anti-unitary symmetry is defined as 
\bea
R(\tilde\bst)=X_LY_R\cdot\mathcal{K}
\eea
As a result, the associated order parameter manifold is 
\bea
\mathcal{M}=G/H=U(1)\simeq S^1
\eea

(3) If the N\'eel vector has both in-plane and $\hat z$ components, both $U(1)_s$ rotation and layer exchange $\mathcal{X}$ are broken, leaving anti-unitary $\bst_{\bf n}$ defined in (\ref{in-plane time reversal}) an unbroken symmetry. Therefore the remnant symmetry group is 
\bea
H(n_z\neq0,n_xn_y\neq0)=O(2)_c\times \mathbb{Z}_2^{\bst_{\bf n}}
\eea
and the associated order parameter manifold is
\bea
\mathcal{M}=G/H=O(2)
\eea

A summary of the long range orders and associated topological defects of the local order parameters is shown in Table \ref{tab:Neel ordered phases}. 

\begin{table}[t]
\begin{tabular}{|c|c|c|c|c|}
\hline
N\'eel vector&Unbroken subgroup $H$&$G/H$&Topological defects&Gauge group\\
\hline
${\bf n}\parallel\hat z$&$O(2)_c\times U(1)_s\rtimes \mathbb{Z}_2^{\mathcal{X}\cdot\bst}$&$\mathbb{Z}_2$&Domain wall&$\mathbb{Z}_2\times\mathbb{Z}_2$\\
\hline
${\bf n}\perp\hat z$&$O(2)_c\times \mathbb{Z}_2\times \mathbb{Z}_2^{\tilde\bst}$&$S^1$&Vortex&$\mathbb{Z}_2$\\
\hline
$n_z\neq0,n_xn_y\neq0$&$O(2)_c\times \mathbb{Z}_2^{\bst_{\bf n}}$&$O(2)$&Domain wall + Vortex&$\mathbb{Z}_2$\\
\hline
\end{tabular}
\caption{A summary of distinct long range orders for different orientations of the N\'eel vector ${\bf n}$, and the associated gauge group for their topological orders.}\label{tab:Neel ordered phases}
\end{table}

\subsection{Local and string order parameters for the long range N\'eel order}

According to Lieb's theorem\cite{Lieb1994}, for the model (\ref{fixed flux ham}) of interacting fermions with bipartite hopping and particle-hole symmetry, a uniform $\pi$ flux per plaquette has the lowest energy. As a result, in the low-energy manifold below the vison gap, we can consider a $\pi$-flux Hubbard model of fermions $\{f_{\nu,j}\}$ on the square lattice. To understand the topological nature of the ground state and its excitations, we need to understand the wavefunctions of the bilayer spin liquid model. 

In the low-energy manifold, since the flux (\ref{flux operator}) in every plaquette is fixed to $F_{\nu,ijkl}\equiv-1,~\forall~\nu$, we can choose a gauge so that 
\bea\label{gauge fix}
\imth b^{\gamma_{ij}}_ib^{\gamma_{ij}}_j=u_{ij},~~~\forall~\nu=1,2
\eea
where $u_{ij}$ is fixed as shown in Fig. \ref{Fig:S1}. In the fixed gauge, let's denote the eigenstates for the Hubbard model (\ref{fixed flux ham}) of fermions $\{f_{\nu,j}\}$ (or equivalently of $\{\psi_j\}$) as $\dket{\psi}$. Then the mean-field ansatz $\dket{MF}$ of all fermions $\{c_{\nu,j}^{x,y},b_{\nu,j}^\gamma\}$ can be written as
\bea
\dket{MF}=\dket{\psi}\otimes\dket{\imth b^{\gamma_{ij}}_ib^{\gamma_{ij}}_j=u_{ij}}
\eea
where $\dket{\imth b^{\gamma_{ij}}_ib^{\gamma_{ij}}_j=u_{ij}}$ denotes the unique eigenstate of fermions $\{b^\gamma_{\nu,j}\}$ that satisfies the fixed gauge (\ref{gauge fix}). The physical eigenstate of the bilayer spin liquid model (\ref{eq:H}) is then obtained by enforcing the onsite constraint (\ref{onsite constraint}) on the mean-field ansatz:
\bea\label{physical eigenstate}
\dket{QSL}=\prod_{j,\nu}\frac{1+D_{\nu,j}}2\dket{MF}
\eea
via the projection operator $\frac{1+D_{\nu,j}}2$ on every site. 

In the N\'eel order of the square-lattice Hubbard model, there is an off-diagonal long range order for the N\'eel vector (\ref{Neel vector}) in the mean-field ansatz:
\bea
\lim_{|i-j|\rightarrow\infty}(-1)^{i_x+i_y+j_x+j_y}\expval{MF|f^\dagger_{\mu,i}\vec\sigma_{\mu\nu}f_{\nu,i}\cdot f^\dagger_{a,j}\vec\sigma_{ab}f_{b,j}|MF}\neq0
\eea
However, this does not guarantee the same local order parameter to exist in the physical ground state (\ref{physical eigenstate}). More precisely, while the $\hat z$-component of the N\'eel vector
\bea\label{local order parameter:z-Neel}
n^z_j\propto\sum_{\mu=1,2}\mu f^\dagger_{\mu,j}f_{\mu,j}=\sum_\nu\kappa\Gamma^{5}_{\nu,j}
\eea
where $\kappa=-1,1$ for $\nu=1,2$ respectively. $n^z_j$ is a local operator and, the in-plane components
\bea\label{spin flip}
n^+_j=n^x_j+\imth n^y_j\propto f^\dagger_{2,j}f_{1,j}
\eea
do not commute with the onsite constraint (\ref{onsite constraint}). As a result, the off-diagonal long range order of local order parameters in the mean-field anstaz will translate into a string order parameter in the physical wavefunction:
\bea\label{string order}
\lim_{|i-j|\rightarrow\infty}(-1)^{i_x+i_y+j_x+j_y}\expval{QSL|f^\dagger_{2,i}f_{1,i}\big(\prod_\nu\prod_{a=1}^M\imth b^{\gamma_{j_a,j_{a+1}}}_{\nu,j_a}b^{\gamma_{j_a,j_{a+1}}}_{\nu,j_{a+1}} \big)f^\dagger_{1,j}f_{2,j}|QSL}\neq0
\eea
where $(j_1,j_2,\cdots,j_M)$ is a string of consecutive sites that connect $j_1=i$ and $j_M=j$. It is straightforward to check that the string order parameter above commutes with the onsite constraint (\ref{onsite constraint}). Although the in-plane component $n^+$ of the N\'eel vector is not a local operator, there is a local order parameter for the in-plane N\'eel order:
\bea\label{local order parameter:in-plane Neel}
S^+_{\expval{ij}}\equiv f^\dagger_{2,i}f_{1,i}(\prod_\nu\imth b^{\gamma_{ij}}_{\nu,i}b^{\gamma_{ij}}_{\nu,j})f^\dagger_{2,j}f_{1,j}
\eea
where $\expval{i,j}$ is a pair of nearest neighbor sites. And one can check that the in-plane N\'eel order in the mean-field ansatz $\dket{MF}$ implies the following long range order of the physical wavefunction $\dket{QSL}$:
\bea
\lim_{|i-i^\prime|\rightarrow\infty}\expval{QSL|S^+_{\expval{ij}}S^-_{\expval{i^\prime j^\prime}}|QSL}\neq0
\eea

\subsection{Topological orders in the N\'eel ordered phase}

For a small value of $J/K$, the $\pi$-flux Hubbard model (\ref{fixed flux ham}) has a gapless spectrum with massless Dirac fermion excitations. Each layer hosts one type of vison (i.e. $\pi$-flux) excitations, giving rise to a $\mbz_2\times\mbz_2$ gauge field. The low energy physics is therefore described by the two decoupled layers of gapless $\mbz_2$ spin liquids, each layer described by two branches of massless Dirac fermions coupled to a $\mbz_2$ gauge field.

In the large $J/K$ case, a long-range N\'eel order develops in the ground states of the $\pi$-flux Hubbard model (\ref{fixed flux ham}), where the Dirac fermions acquires a mass, giving rise to a gapped spectrum for all fermions. Different orientations of the N\'eel vector (\ref{Neel vector}) not only give rise to distinct long range orders with different order parameters, but also distinct topological orders with different gauge groups. Below we analyze the 3 different orientations of the N\'eel vector summarized in Table \ref{tab:Neel ordered phases}. 
\subsubsection{${\bf n}\parallel\hat z$}

As previously discussed, when the N\'eel vector aligns along $\hat z$-axis, the $\hat z$-component of the N\'eel vector is a local order parameter for the Ising-type N\'eel order, with a order parameter manifold of $G/H=\mathbb{Z}_2$. Since the fermions are all gapped, the ground state exhibits a $\mbz_2\times\mbz_2$ topological order, due to deconfined vison excitations from each layer. 

The only gapless excitations in this case are the Goldstone modes of the N\'eel order, described by the spin flip operator in (\ref{spin flip}). However, the operator $n^+_j\sim(-1)^{j_x+j_y}f^\dagger_{2,j}f_{1,j}$ is not a gauge invariant operator as it violates the onsite constraint (\ref{onsite constraint}). The gapless Goldstone modes of the Hubbard model lead to a power-law decaying correlation of the string order parameter (\ref{string order}) in this case.

If the system is gapped, it will naively be an Abelian $\mathbb{Z}_2\times \mathbb{Z}_2$ topological order described by a 4-component Chern-Simons theory with the following ${\bf K}$ matrix\
\bea
{\bf K}=\bpm0&2\\2&0\epm\oplus\bpm0&2\\2&0\epm
\eea
Since both $U(1)_c$ and $U(1)_s$ symmetries are preserved, they are captured by charge and pseudo-spin vectors
\bea
{\bf q}_c=(2,0,2,0)^T,~~~{\bf q}_s=(2,0,-2,0)^T.
\eea

However, the nature of the gapless ``Goldstone modes'' here, plays a crucial role to interpret the physical ground state (\ref{physical eigenstate}) after the projection $\frac{1+\hat D_{\nu,j}}2$ on each site. Notice, the two real Goldstone modes together form a complex bosonic mode
\bea\label{algebraic spin liquid}
n^x+\imth n^y\sim b_{\vec k=(\pi,\pi)}\sim\sum_i(-1)^{i_x+i_y}f^\dagger_{2,i}f_{1,i}
\eea
which is gapless, with a linear dispersion. This mode carries a unit gauge charge of both $\mathbb{Z}_2$ gauge fields, one from each layer, and is hence not a local excitation. Therefore, this ground state should be understood as an algebraic $\mbz_2\times \mbz_2$ spin liquid, where the bound state (\ref{algebraic spin liquid}) (with quasiparticle vector $(1,1,-1,1)^T$ in the Abelian Chern-Simons theory) of two types of fermionic spinons $f_{2}$ (with quasiparticle vector $(1,1,0,0)^T$) and $f_{1}$ (with quasiparticle vector $(0,0,-1,1)^T$) becomes gapless with an algebraic correlation in (\ref{string order}). The mode $b$ obeys bosonic self statistics, but is really an anyon excitation with semionic mutual braiding with fermions $f_{\nu,j}$ from either layer. Therefore the system exhibits fractionalized Goldstone modes that obey anyonic statistics.

After incorporating the gapless anyon $b$ in (\ref{algebraic spin liquid}) into the low energy description, the effective field theory for this algebraic spin liquid writes
\bea
\mathcal{L}_\text{ASL}=\sum_{I,J}\frac{\epsilon^{\mu\nu\rho}}{4\pi}a_\mu^I{\bf K}_{I,J}\partial_\nu a_\rho^J-\sum_{\alpha=c,s}\sum_I\frac{\epsilon^{\mu\nu\rho}}{2\pi}A^\alpha_\mu{\bf q}^\alpha_I\partial_\nu a_\rho^I+
|(-\imth\partial_\mu-2A^s_\mu-a^1_\mu-a^2_\mu+a^3_\mu-a^4_\mu)^2b|^2+\cdots
\eea
where $b$ denotes the critical anyon mode out of the Goldstone modes of the broken emergent symmetry. Typically, the robust presence of a gapless anyon excitation in a system requires a mechanism to protect it. In this case, it is the emergent pseudo-spin $SU(2)_s$ symmetry that protects the criticality of this anyon mode.

\subsubsection{${\bf n}\perp\hat z$}
It's straightforward to show that a nonzero in-plane N\'eel component $\expval{b}\neq0$ corresponds to condensing the anyon characterized by vector $(1,1,-1,1)$ in the Chern-Simons theory, and drives the system into a usual $\mbz_2$ topological order described by a 2-component Abelian Chern-Simons theory with ${\bf K}=\bpm0&2\\2&0\epm$. 

This can be understood by considering the topological defects of the local order parameter (\ref{local order parameter:in-plane Neel}): since the order parameter manifold for an in-plane N\'eel order (${\bf n}\perp\hat z$) is $G/H=U(1)\simeq S^1$, the long-range order supports point defects (i.e. vortices) classified by an integer-valued vorticity $\pi_1(G/H)=\mbz$. The vortex of local order parameter $S^+=S^x+\imth S^y$ in (\ref{local order parameter:in-plane Neel}) is nothing but a $\pi$ flux from either layer, characterized by quasiparticle vector $(1,0,0,0)$ or $(0,0,1,0)$ in the 4-component Abelian Chern-Simons theory. Since the vortices are logarithmically confined in two spatial dimensions, the $\pi$ flux from either layer is confined with an in-plane N\'eel order. As a result, the ground state features a $\mbz_2$ topological order, whose vison excitation is the bound state of a $\pi$ flux from both layers. 

Next we describe the Goldstone modes in the in-plane N\'eel order. Without loss of generality, we assume the N\'eel vector to point along $\hat x$ axis in the ground state, i.e. $\expval{n^x}\neq0$ in the ground state of the gauge-fixed $\pi$-flux Hubbard model (\ref{fixed flux ham}). In this case, we can write down the following local operator that creates the Goldstone mode $n^y$:
\bea
\mathcal{N}^y=\sum_in_i^x\prod_\nu (\imth b^\gamma_{\nu,i}b^\gamma_{\nu,i+\hat e_\gamma})n^y_{i+\hat e_\gamma}
\eea
where $\hat e_\gamma$ is the unit lattice vector along bond $\gamma$ direction. Due to the long-range string order (\ref{string order}) in the ground state, it is straightforward to show that
\bea
\mathcal{N}^y\dket{QSL}\sim(\prod_{\nu,i}\frac{1+D_{\nu,i}}2)n^y\dket{MF}
\eea

\subsubsection{$n_z\neq0,~n_xn_y\neq0$}
When the N\'eel vector has both in-plane and $\hat z$ components, the ground state also exhibits a $\mbz_2$ topological order, for the same reason as discussed above. With an order parameter manifold of $G/H=O(2)$, the topological defects in this case include both domain walls classified by $\pi_1(O(2))=\mathbb{Z}_2$, and vortices classified by $\pi_1(O(2))=Z$. 

\section{Effective $\eta-\rho$ Hamiltonian in the large-$J$ limit}

In this section, we provide the details of the derivation for the effective Hamiltonian to fourth-order in $|K/J|\ll 1$, which we label as an ``$\eta-\rho$" model. For $K=0$, the ground state (GS) is doubly degenerate, $\ket{\uparrow_1\uparrow_2},\ket{\downarrow_1\downarrow_2}$ at each site. These doublets can be represented by the pseudospins $\ket{\Uparrow},\ket{\Downarrow}$ with associated operators
\begin{align}
\eta^z_{i} =&\frac{1}{4}\left(\sigma_{1i}^z+\sigma_{2i}^z \right) \notag \\
\eta^\pm_{i} = & \frac{1}{4} \sigma_{1i}^{\pm}\sigma_{2i}^{\pm} 
\end{align}
which obey an SU(2) algebra. Note that $\eta^z_{i}$ is a dipolar operator whereas $\eta^x_{i}$ and $\eta^y_{i}$,
\begin{align}
\eta^x_{i}=& \frac{1}{4}\left(\sigma_{1i}^x\sigma_{2i}^x-\sigma_{1i}^y\sigma_{2i}^y \right) \nonumber
\\
\eta^y_{i}=& \frac{1}{4}(\sigma_{1i}^x\sigma_{2i}^y+\sigma_{1i}^y\sigma_{2i}^x)
\label{Eq:EN_eta}
\end{align}
are quadrupole operators. Apart from the pseudospin doublet, there is an additional four-fold degeneracy per site, arising from the orbital DOF. We introduce an operator 

\noindent \begin{eqnarray}
P_0=\prod_i\frac{1+\sigma^z_{1i}\sigma^z_{2i}}{2},
\end{eqnarray}

\noindent which projects out the states $\ket{\uparrow_1\downarrow_2},\ket{\downarrow_1\uparrow_2}$ at each site.

In the following, we derive the effective Hamiltonian perturbatively to fourth-order in $K$. We also note that all odd-ordered terms vanish since $H_K$ only connects states inside the GS manifold to states outside of it. 

\subsection{Second-order effective Hamiltonian $H_{g_{2}}$ }

The effective Hamiltonian to second order is 
\begin{eqnarray}
H^{(2)}=P_0H_KSH_KP_0
\label{eq:p2}
\end{eqnarray}
where
\begin{eqnarray}
S=\frac{(1-P_0)}{(E_0-H_J)},
\end{eqnarray}

\noindent and $H_{J}$ is the inter-layer coupling. In order to evaluate eq.~\ref{eq:p2}, we introduce the operators 
\begin{eqnarray}
X_{ij}=\sigma_{1i}^+\sigma_{1j}^-\tau_{1i}^\gamma\tau_{1j}^\gamma+\sigma_{2i}^-\sigma_{2j}^+\tau_{2i}^\gamma\tau_{2j}^\gamma
\end{eqnarray}
and re-write $H_K= \sum_{\langle ij \rangle_\gamma}X_{ij}+X_{ji}$. Since $P_0X_{ij}^2P_0=0$, we obtain  
\begin{eqnarray}
H^{(2)}
= \frac{K^2}{4J}\sum_{\langle ij \rangle_\gamma}P_0\{X_{ij},X_{ji}\}P_0.
\label{Eq:EN_H2}
\end{eqnarray}
Furthermore, we use the identities
\begin{eqnarray}
P_0\sigma_{1i}^+\sigma_{2i}^+P_0&=&4\eta_i^+ \nonumber \\
P_0\sigma_{1i}^-\sigma_{2i}^-P_0&=&4\eta_i^- \nonumber \\
P_0\sigma_{\nu i}^+\sigma_{\nu i}^-P_0&=&2(1+2\eta_i^z) 
\nonumber \\
P_0\sigma_{\nu i}^-\sigma_{\nu i}^+P_0&=&2(1-2\eta_i^z),
\label{eq:eta}
\end{eqnarray}

\noindent to express 

\noindent \begin{equation}
    H^{(2)}=-g_2 \sum_{\langle ij \rangle_\gamma} \left[ -\eta_i^z\eta_j^z+ \frac{1}{2}(\eta^+_i\eta^-_j+\eta^-_i\eta^+_j) \tau_{1i}^{\gamma}\tau_{1j}^{\gamma}\tau_{2i}^{\gamma}\tau_{2j}^{\gamma} \right],
\end{equation}

\noindent where $g_2=K^2/4J$. We perform  rotations by $\pi$ about the $z$-axis on the $\eta$ pseudo-spins at every other site to obtain the effective Hamiltonian at $g_{2}$ level

\noindent \begin{equation}
    H_{g_2}=g_2 \sum_{\langle ij \rangle_\gamma} \left[ \eta_i^z\eta_j^z+ \frac{1}{2}(\eta^+_i\eta^-_j+\eta^-_i\eta^+_j) \tau_{1i}^{\gamma}\tau_{1j}^{\gamma}\tau_{2i}^{\gamma}\tau_{2j}^{\gamma} \right]
    \label{Eq:EN_Hg2}.
\end{equation}

\noindent Note that $\tau_{1i}^{\gamma}\tau_{1j}^{\gamma}\tau_{2i}^{\gamma}\tau_{2j}^{\gamma} = \rho^{\gamma}_{ij}$, where $\rho^{\gamma}_{ij}= q^{\gamma}_{i} q^{\gamma}_{j}$.

\subsection{Fourth-order effective Hamiltonian $H_{g_{4}}$ }

The general expression to fourth order reads
\begin{eqnarray}
H^{(4)}=P_0H_KSH_KSH_KSH_KP_0- \frac{1}{2}\left(P_0H_KS^2H_KP_0H_KSH_KP_0 + P_0H_KSH_KP_0H_KS^2H_KP_0 \right)
\label{eq:p4}
\end{eqnarray}

\noindent We focus on the first term, which, as described below, includes coupled pseudo-spin and orbital DOF around the $p/p'$ plaquettes. These play a critical role in lifting the extensive degeneracy of the GS of the leading $H_{g_{2}}$, due to the orbital DOF, as implied by $\rho^{\gamma}_{ij}=1~\forall~\gamma, \braket{ij}$ (Supplementary section~\ref{Sec:Hg2}). By contrast, the remaining terms in eq.~\ref{eq:p4} are either independent of the orbital DOF or involve two pairs of NN pseudo-spins coupled to $\rho^{\gamma}_{ij} \rho^{\gamma'}_{jk}$, where $\gamma \neq \gamma'$ for the two connected bonds. These involve subleading corrections to the pseudo-spin configuration, but do not lift the GS degeneracy of $H_{g_{2}}$. Consequently, we ignore these contributions and focus instead on the plaquette terms in the following. To illustrate, we consider a single $ p$ plaquette (Supplementary figure~\ref{Fig:S1}) in the first term as 
\begin{align}
    P_0H_KSH_KSH_KSH_KP_0 
    =\frac{K^4}{8J^3}&P_0[-2\{X_{ij}X_{kl},X_{jk}X_{li}\}-\{\{X_{ij},X_{jk}\},\{X_{kl},X_{li}\}\} \notag\\
    & -\{\{X_{jk},X_{kl}\},\{X_{li},X_{ij}\}\}]P_0 
    % \nonumber \\
    % =\frac{K^4}{8J^3}&P_0[-10\{X_{ij}X_{kl},X_{jk}X_{li}\}-3\{X_{ik},X_{ki}\}-3\{X_{jl},X_{lj}\}+ 2\{X_{kl},X_{lk}\} \nonumber \\
    %&\qquad \qquad \qquad \ \ 2 
    % & + 2\{X_{li},X_{il}\}+2\{X_{ij},X_{ji}\}+2\{X_{jk},X_{kj}\}]P_0 + 
    + \hdots
    \label{Eq:Frst_trm}
\end{align}

\noindent %Note that terms involving products of two identical $X_{ij}$ operators vanish under projection by $P_{0}$. 
where $\hdots$ includes all remaining $p/p'$ plaquettes. We simplify the expressions by using the identities in eq. \ref{eq:eta} and by rotating the pseudo-spins such that $\eta_i^\pm\rightarrow-\eta_i^\pm$ at every other site. Carrying out similar operations for the remaining plaquettes, we obtain the effective Hamiltonian at fourth order

\begin{align}
    H_{g_4}=g_4\sum_p[\eta^\square_{ijkn}(W_{1p}+W_{2p})+\eta^\square_{klmn}(W_{1p^\prime}+W_{2p^\prime})],
    \label{eq:Hg4}
\end{align}

\noindent where

\begin{align*}
   \eta^\square_{ijkn}&=[5\eta^z_i\eta^z_j\eta^z_k\eta^z_n+5/2(\eta_i^+\eta_j^-\eta_k^+\eta_n^- +\eta_i^-\eta_j^+\eta_k^-\eta_n^+)\\
    &-5/2(\eta^+_i\eta^-_j\eta^z_k\eta^z_n\ \rho_{ij} \rm + 11 \ permutations)+ \\
    &[(\eta_i^z\eta_j^z+1/2(\eta_i^+\eta_j^-+\eta_i^-\eta_j^+)\ \rho_{ij}) +\\ &(\eta_j^z\eta_k^z+1/2(\eta_j^+\eta_k^-+\eta_j^-\eta_k^+)\ \rho_{jk})+\\
    &(\eta_k^z\eta_n^z+1/2(\eta_k^+\eta_n^-+\eta_k^-\eta_n^+))+\\
    &(\eta_n^z\eta_i^z+1/2(\eta_n^+\eta_i^-+\eta_n^-\eta_i^+)\ \rho_{ij})+\\
    &(\eta_i^z\eta_k^z+1/2(\eta_i^+\eta_k^-+\eta_i^-\eta_k^+)\ \rho_{ni})+\\
    &(\eta_j^z\eta_n^z+1/2(\eta_j^+\eta_n^-+\eta_j^-\eta_n^+)\ \rho_{jk})+1]].
\end{align*}

\noindent Note that these are due to the terms in eq.~\ref{Eq:Frst_trm}.

\section{The effects of $\tau_{\nu, i}$ and $W_{\nu p/p'}$ operators in the $\eta-\rho$ basis}

We first consider $\tau_{\nu, i}$ operating on any local orbital state determined by the set $\{q^{x}_{i}, q^{y}_{i}, q^{z}_{i} \}$  subject to the constraint $q^{x}_{i} q^{y}_{i} q^{z}_{i} =-1$:

\begin{eqnarray}
\tau_1^{x} |-,-,-\rangle &= &- |-,+,+\rangle \\ \nonumber
\tau_2^{x} |-,-,-\rangle &= &+ |-,+,+\rangle \\ \nonumber
\tau_1^{y} |-,-,-\rangle &= &i |+,-,+\rangle \\ \nonumber
\tau_2^{y} |-,-,-\rangle &= &-i |+,-,+\rangle \\ \nonumber
\tau_1^{z} |-,-,-\rangle &= &+ |+,+,-\rangle \\ \nonumber
\tau_2^{z} |-,-,-\rangle &= &- |+,+,-\rangle \\ \nonumber
\tau_1^{x} |-,+,+\rangle &= &- |-,-,-\rangle \\ \nonumber
\tau_2^{x} |-,+,+\rangle &= &+ |-,-,-\rangle \\ \nonumber
\tau_1^{y} |-,+,+\rangle &= &i |+,+,-\rangle \\ \nonumber
\tau_2^{y} |-,+,+\rangle &= &i |+,+,-\rangle \\ \nonumber
\tau_1^{z} |-,+,+\rangle &= &+ |+,-,+\rangle \\ \nonumber
\tau_2^{z} |-,+,+\rangle &= &+ |+,-,+\rangle \\ \nonumber
\tau_1^{x} |+,-,+\rangle &= &+ |+,+,-\rangle \\ \nonumber
\tau_2^{x} |+,-,+\rangle &= &+ |+,+,-\rangle \\ \nonumber
\tau_1^{y} |+,-,+\rangle &= &-i |-,-,-\rangle \\ \nonumber
\tau_2^{y} |+,-,+\rangle &= &+i |-,-,-\rangle \\ \nonumber
\tau_1^{z} |+,-,+\rangle &= &+ |-,+,+\rangle \\ \nonumber
\tau_2^{z} |+,-,+\rangle &= &+ |-,+,+\rangle \\ \nonumber
\tau_1^{x} |+,+,-\rangle &= &+ |+,-,+\rangle \\ \nonumber
\tau_2^{x} |+,+,-\rangle &= &+ |+,-,+\rangle \\ \nonumber
\tau_1^{y} |+,+,-\rangle &= &-i |-,+,+\rangle \\ \nonumber
\tau_2^{y} |+,+,-\rangle &= &-i |-,+,+\rangle \\ \nonumber
\tau_1^{z} |+,+,-\rangle &= &+ |-,-,-\rangle \\ \nonumber
\tau_2^{z} |+,+,-\rangle &= &+ |-,-,-\rangle 
\label{Eq:tau}
\end{eqnarray}

\noindent In summary, $\tau_{1/2,i}^{x}$  preserve $q^{x}_{i}$ while changing the signs of $q^{y/z}_{i}$, and so on for all permutations. In terms of the $\rho^{\gamma}_{ij} = q^{\gamma}_{i}q^{\gamma}_{j}$ bonds , the same operator changes the signs of $\rho^{y}_{ij}$ and $\rho^{z}_{ik}$, where $ij$ and $ik$ are NN pairs.

The flux operators are defined by 

\noindent \begin{align}
W_{\nu p}= & \sigma^z_{\nu k}\sigma^z_{\nu n}\tau^x_{\nu i}\tau^y_{\nu j}\tau^x_{\nu k}\tau^y_{\nu n} \\ 
W_{\nu p'}= & \sigma^z_{\nu k}\sigma^z_{\nu n}\tau^x_{\nu l}\tau^y_{\nu m}\tau^x_{\nu k}\tau^y_{\nu n},
\end{align} 

\noindent where the index convention is illustrated in Supplementary figure~\ref{Fig:S1}. These operators act both on the spin ($\sigma$), as well as on the orbital ($\tau$) DOF. The labels $p,p'$ label the plaquettes, as illustrated in Supplementary figure~\ref{Fig:S1}. By applying eq.~\ref{Eq:tau} around a single $p$ plaquette, we see that $W_{1p}$ flips the $q^{\gamma}$'s at four of the six sites of the unit cell. Consequently, six $\rho^{\gamma}_{ij}$ bonds change signs, as shown in Supplementary figure ~\ref{Fig:S2}~(a). The effect of $W_{1p'}$ is similar (Supplementary figure~\ref{Fig:S2}~(b)) although the end state differs in the $q^{\gamma}$ values at each of the six sites. Due to the presence of the $\sigma^{z}_{k}$ and $\sigma^{z}_{n}$ operators, both $W_{1 p}$ and $W_{1p'}$ also rotate the in-plane pseudo-spin components at sites $k$ and $n$ as $\eta^{x,y}_{k} \rightarrow -\eta^{x,y}_{k}$ and $\eta^{x,y}_{n} \rightarrow - \eta^{x,y}_{n}$. 

\begin{suppfigure}[t]
\includegraphics[width=10 cm]{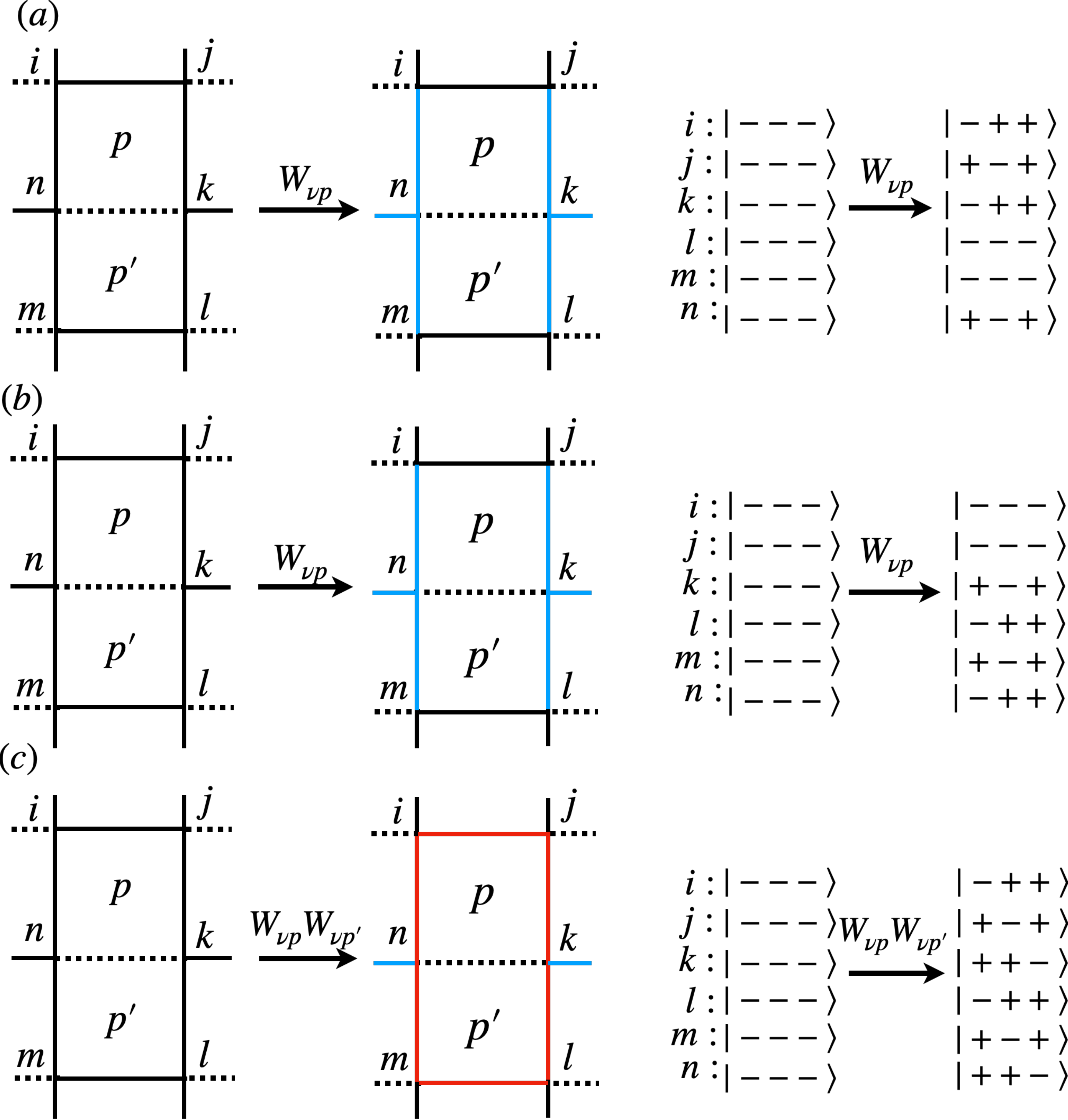}
\caption{The effects of the flux operators on the $q^{\gamma}_{i}$ orbital states and $\rho^{\gamma}_{ij}$ bond configurations. Black lines indicate positive bonds with all $q^{\gamma}_{i}=-1$ around the unit cell. Blue lines indicate negative bonds, while red lines are positive bonds with non-trivial $q^{\gamma}_{i}$ values. The right panels show how the latter transform. We illustrate for (a) $W_{1p}$, (b)$W_{1p'}$, and (c) $W_{1p}W_{1p'}$.
}
\label{Fig:S2}
\end{suppfigure}

\section{Exact diagonalization of $H_{g_{2}}$}

\label{Sec:Hg2}

\begin{suppfigure}[t]
\includegraphics[width=\columnwidth]{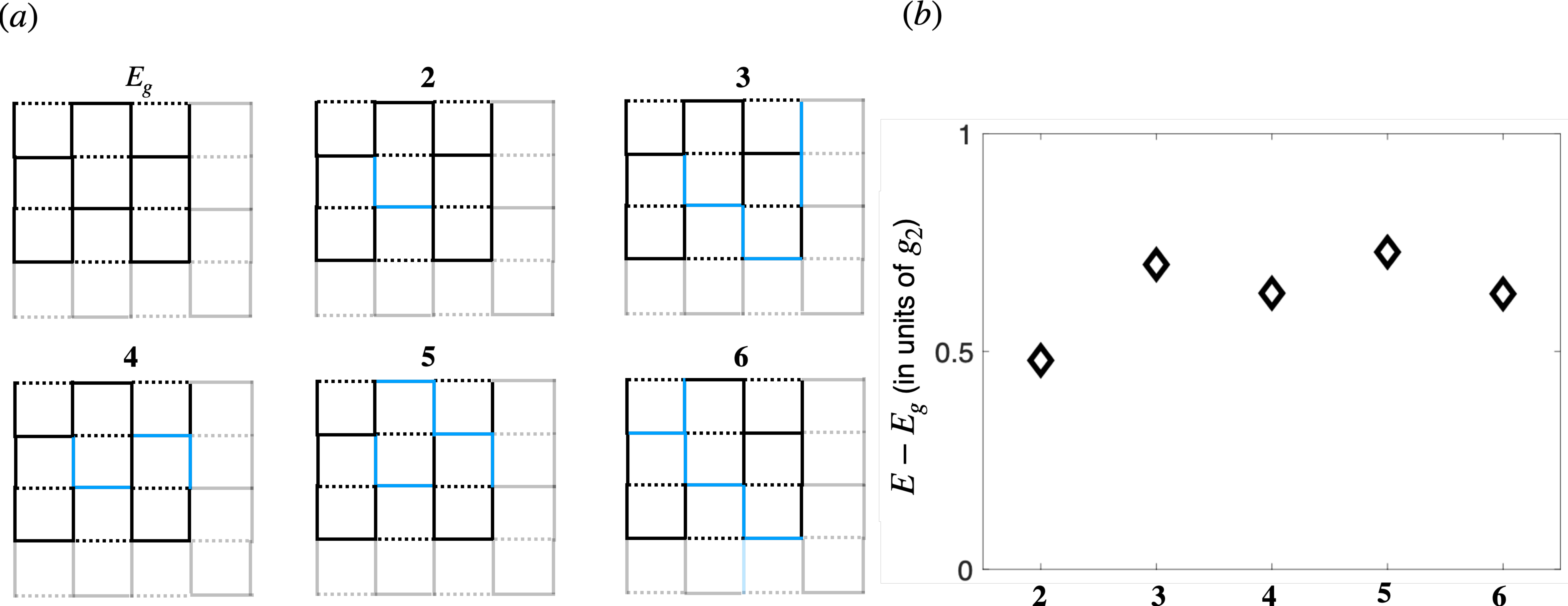}
\caption{Bond configurations and ground-state energy for a $4\times 4$ lattice on a torus. (a) Dotted bonds correspond to the identity bond while gray bonds illustrate wrapping around the torus. The configurations shown correspond to (1) Uniform $\rho^{\gamma}_{ij}=1$, and (2-6) strings of negative bonds. (b) Difference in the ground-state energies corresponding to (a). The uniform configuration is lowest in energy.}
\label{Fig:S3}
\end{suppfigure}

The effective Hamiltonian 

\begin{align}
    H_{g_2}= & g_2  \bigg[ \sum_{\langle ij \rangle}\eta_i^z\eta_j^z+ \sum_{\langle ij \rangle_{\gamma} } (\eta^x_i\eta^x_j+\eta^y_i\eta^y_j)\rho^\gamma_{ij} \bigg] + \sum_{i} (-1)^{i_{x}+i_{y}} \left( h_{x} \eta^{x}_{i} + h_{z} \eta^{z}_{i} \right),
\end{align}

\noindent couples the in-plane pseudo-spin components $\eta^{x/y}_{i}$ with the orbital bonds $\rho^{\gamma}_{ij}= q^{\gamma}_{i} q^{\gamma}_{j}$. The latter are preserved and $H_{g_{2}}$ can be solved for the pseudo-spins in any given bond configuration which is consistent with the local constraint $\prod_{\gamma} q^{\gamma}_{i}=-1$. Here, we consider $H_{g_{2}}$ on a $4 \times 4$ torus with both $h_{x/z}=0$. We determine the GS energy via exact diagonalization in the pseudo-spin sector for various bond configurations. We find that the uniform $\rho^{\gamma}_{ij}=1$ configuration has the lowest energy. In this sector, $H_{g_{2}}$ maps onto a 2D Heisenberg model with periodic boundary conditions. In contrast, all states with one or more strings of $\rho^{\gamma}_{ij}=-1$ bonds have higher energy, as shown in Supplementary figure~\ref{Fig:S3}. 

Note that exceptions to these conclusions occur for $h_{z} > 0, h_{x}=0$, or when the pseudo-spin spontaneously order along the $z$ direction only, in the infinite-size system. As shown in Supplementary section~\ref{Sec:Strn}, a non-contractible string of $\rho^{\gamma}_{ij}=-1$ bonds becomes degenerate with the uniform $\rho^{\gamma}_{ij}=1$ state.   

\section{Exact diagonalization of $H_{g_{4}}$}

As discussed in Supplementary section~\ref{Sec:Hg2}, a state with uniform $\rho^{\gamma}_{ij}=1$ bonds and zero net pseudo-spin, as in a 2D Heisenberg model, has the lowest energy at $H_{g_{2}}$ level. For $h_{x}=0$, $H_{g_{2}}$ commutes with the flux operators $W_{\nu p/p'}$. Therefore, the states

\noindent \begin{align}
\prod_{\{\nu, p\}, \{\nu', p'\}} 
W_{\nu p} W_{\nu' p'} \ket{\eta_{\rm{Heisenberg}},~\forall~\rho^{\gamma}_{ij}=1},
\end{align}   

\noindent where the ket is the GS at $H_{g_{2}}$ level, are degenerate with the latter. The products can include any combinations of $W_{\nu p/p'}$ operators and this implies an extensive degeneracy of the GS. For $h_{x}>0$, $H_{g_{2}}$ only commutes with $W_{\nu, p} W_{\nu, p'}$, where $p/p'$ are in the same unit cell, such that the degeneracy is partially lifted. We shall assume a small $h_{x}$ in the following and ignore this additional complication. 

$H_{g_{4}}$ lifts the extensive degeneracy. To see how, we first project onto states of definite flux as

\noindent \begin{widetext}
\noindent \begin{align}
  \ket{\eta; \{a_{\nu p}, a_{\nu p'} \}} = & \prod_{\nu, p,p'} \frac{1}{2^4}\left(1+a_{\nu p}W_{\nu p}  \right)
\left(1+a_{\nu p'}W_{\nu p'}  \right) \ket{\eta_{\rm{Heisenberg}},~\forall~\rho^{\gamma}_{ij}=1} ~a_{\nu p/p'}= \pm 1. 
\end{align}
\end{widetext}

\noindent $\ket{\eta; \{a_{\nu p}, a_{\nu p'} \}}$ are eigenstates of $W_{\nu p/p'}$ with eigenvalue $a_{\nu p}$, corresponding to a $\pi(1-a_{\nu  p})/2$ flux per plaquette $p$ in layer $\nu$, and similarly for $p'$. We can consider the effect of $H_{g_{4}}$ on the degenerate GS manifold of $H_{g_{2}}$. The former commutes with the fluxes and its matrix elements are diagonal in the projected basis:

\noindent \begin{widetext} \begin{align}
 \braket{\eta; \{a_{\nu p}, a_{\nu p'} \}| H_{g_{4}} | \eta; \{b_{\nu p}, b_{\nu p'} \}}  
 =   \frac{g_4}{2^8}\left[\sum_{\nu,p} \braket{\eta^{\square}_{p}} a_{\nu p} 
+ \sum_{\nu,p'} \braket{\eta^{\square}_{p'}} a_{\nu p'} \right] \delta_{\{a_{\nu p} \}, \{ b_{\nu p}\}} \delta_{\{a_{\nu p'} \}, \{ b_{\nu p'}\}}, 
\end{align} 
\end{widetext}

\noindent where 

\noindent \begin{widetext}
\noindent \begin{align}
   \braket{\eta^{\square}_{p}} = 
   \langle &5(\pmb{\eta}_i\cdot\pmb{\eta}_j)(\pmb{\eta}_k\cdot\pmb{\eta}_n)+5(\pmb{\eta}_i\cdot\pmb{\eta}_n)(\pmb{\eta}_j\cdot\pmb{\eta}_k)-5(\pmb{\eta}_i\cdot\pmb{\eta}_k)(\pmb{\eta}_j\cdot\pmb{\eta}_n)-
    (\pmb{\eta}_i\cdot\pmb{\eta}_j)-
    (\pmb{\eta}_j\cdot\pmb{\eta}_k)-(\pmb{\eta}_k\cdot\pmb{\eta}_n)-(\pmb{\eta}_i\cdot\pmb{\eta}_n) \notag \\
    & -(\pmb{\eta}_i\cdot\pmb{\eta}_k)-(\pmb{\eta}_j\cdot\pmb{\eta}_n)+1 \rangle _{\eta_{\rm{Heisenberg}}}. 
\end{align}
\end{widetext}

\noindent This expression is the expectation value of pseudo-spin ring- and pair-exchange terms on plaquette $p$, in the GS of the 2D Heisenberg model. A similar expression holds for the $p'$ plaquettes. Depending on the sign of the expectation value, $H_{g_{4}}$ lifts the extensive degeneracy at $H_{g_{2}}$ level to leading order, and selects a GS of uniform flux. 
\begin{suppfigure}[t]
\includegraphics[width=\columnwidth]{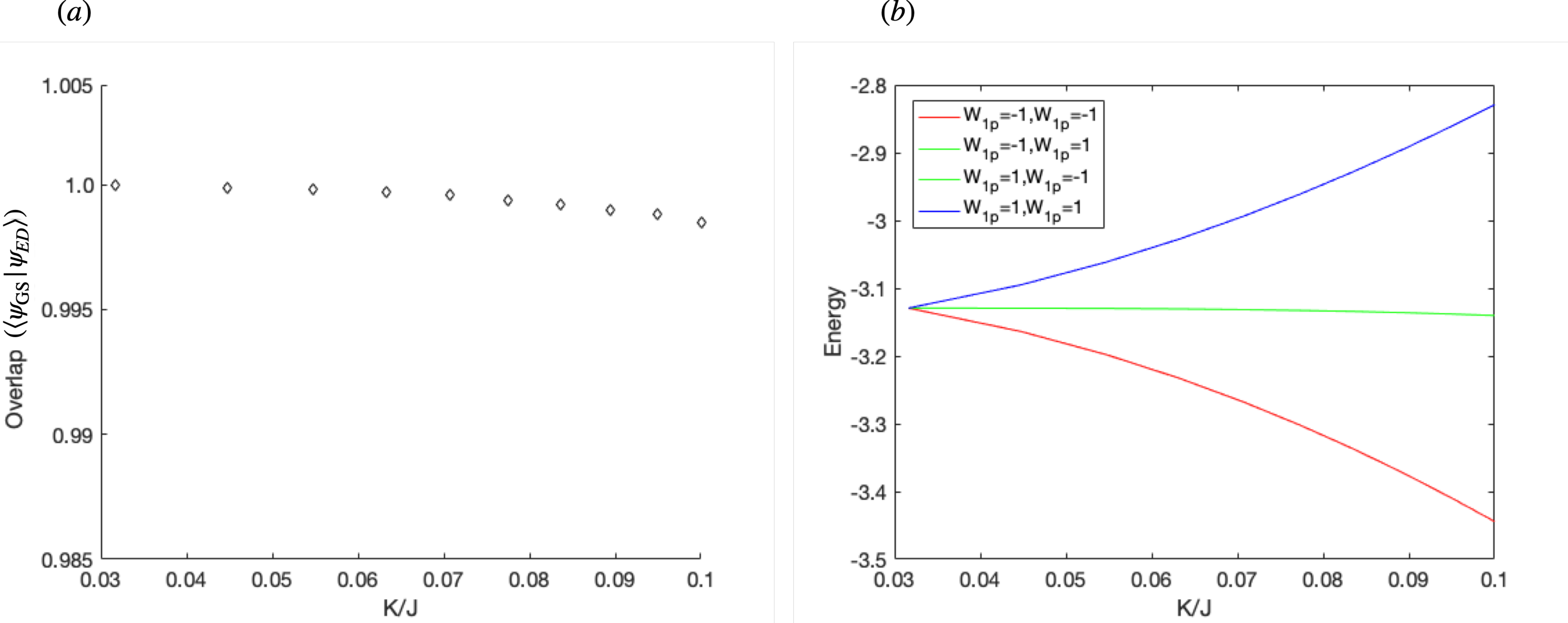}
\caption{(a) The overlap between $\ket{\psi_{\rm GS}}$ and the wavefunction obtained from ED as a fuction of $K/J$. The overlap decreases slowly as the perturbation gets stronger. (b) The lowest energy eigenstate is the state with both $W_{1p}=-1$ and $W_{1p'}=-1$ up to $K/J=0.1$. } 
\label{Fig:S4}
\end{suppfigure}
We confirmed the previous conclusion via exact diagonalizion of $H_{\eta-\rho}$ in a unit cell. The Hilbert space includes all pseudo-spins for uniform $\rho^{\gamma}_{ij}=1$ bonds and all states obtained by applying $W_{\nu p/p'}$ on the latter. The resulting Hamiltonian in the pseudo-spin basis was solved numerically. We find that the GS is unique and corresponds to a state of uniform $\pi$ flux for both $p/p'$ plaquettes (Supplementary figure~\ref{Fig:S4}(b)). We also find that the GS overlaps significantly with $\ket{\eta; \{a_{\nu p}, a_{\nu p'} \}}$ for $g_{4} \ll 
g_{2}$ as shown in Supplementary figure\ref{Fig:S4}(a). 

\section{$\rho^{\gamma}_{ij}$ string defects}
\label{Sec:Strn}

In this section, we consider 

\begin{align}
    H_{g_2}= & \sum_{i} (-1)^{i_{x}+i_{y}} \left( h_{x} \eta^{x}_{i} + h_{z} \eta^{z}_{i} \right) + g_2  \bigg[ \sum_{\langle ij \rangle}\eta_i^z\eta_j^z+ \sum_{\langle ij \rangle_{\gamma} } (\eta^x_i\eta^x_j+\eta^y_i\eta^y_j)\rho^\gamma_{ij} \bigg],
\end{align}

\noindent where $\gamma \in \{x,y,z, I \}$, on a torus. We distinguish two regimes for a) $h_{x} \gg g_{2}, h_{z}=0$ and b) $h_{z} \gg g_{2}, h_{x}=0$, respectively. $H_{g_{2}}$ preserves $\rho^{\gamma}_{ij}$ bonds and can be solved separately in any configuration of the latter, which is consistent with the local constraint $\prod_{i} q^{\gamma}_{i}=-1$. Here, we consider the uniform $\rho^{\gamma}_{ij}=1$ configuration, together with related states which include non-contractible strings of $n$ bonds along which $\rho^{\gamma}_{ij}=-1$. We present a semi-classical analysis and perturbation theory results, which show that the string has infinite energy cost for a), while for b) it vanishes, in the infinite-size system. 

We first proceed by singling out $n$, $\rho^{\gamma}_{ij}$ bonds along a non-contractible loop $C^{x/y}$ which winds once around the torus (Supplementary figure~\ref{Fig:S5}). We re-write the Hamiltonian as 

\noindent \begin{align}
H_{g_2} = H_{\rm{String}} + H_{\rm{Background}} + H_{\rm{Coupling}},
\end{align}

\noindent where $H_{\rm{String}}$ includes all on-site and NN exchange terms  along the path. For the latter, we consider two configurations with $\rho^{\gamma}_{ij}$  bonds either all negative or all positive. Similarly, $H_{\rm{Background}}$ includes the remaining sites, with bonds trivially equal to $1$. $H_{\rm{Coupling}}$ includes all of the exchange terms which connect string and background sites, with associated $\rho^{\gamma}_{ij}=1$ bonds. We calculate corrections to the GS energy in perturbation theory for the $g_{2}$ terms. 

The string of negative bonds around $C^{x/y}$ loops are equivalent to periodic boundary conditions (PBC) for the pseudo-spins along the direction of the loops. By contrast, the boundary conditions in the direction which crosses $C^{x/y}$ are

\noindent \begin{align}
\eta^{x/y}_{i} = & - \eta^{x/y}_{i+N_{2/1}} \\
\eta^{z}_{i} = & \eta^{z}_{i+N_{2/1}}.
\end{align}
\begin{suppfigure}[t]
\includegraphics[width=8.0 cm]
{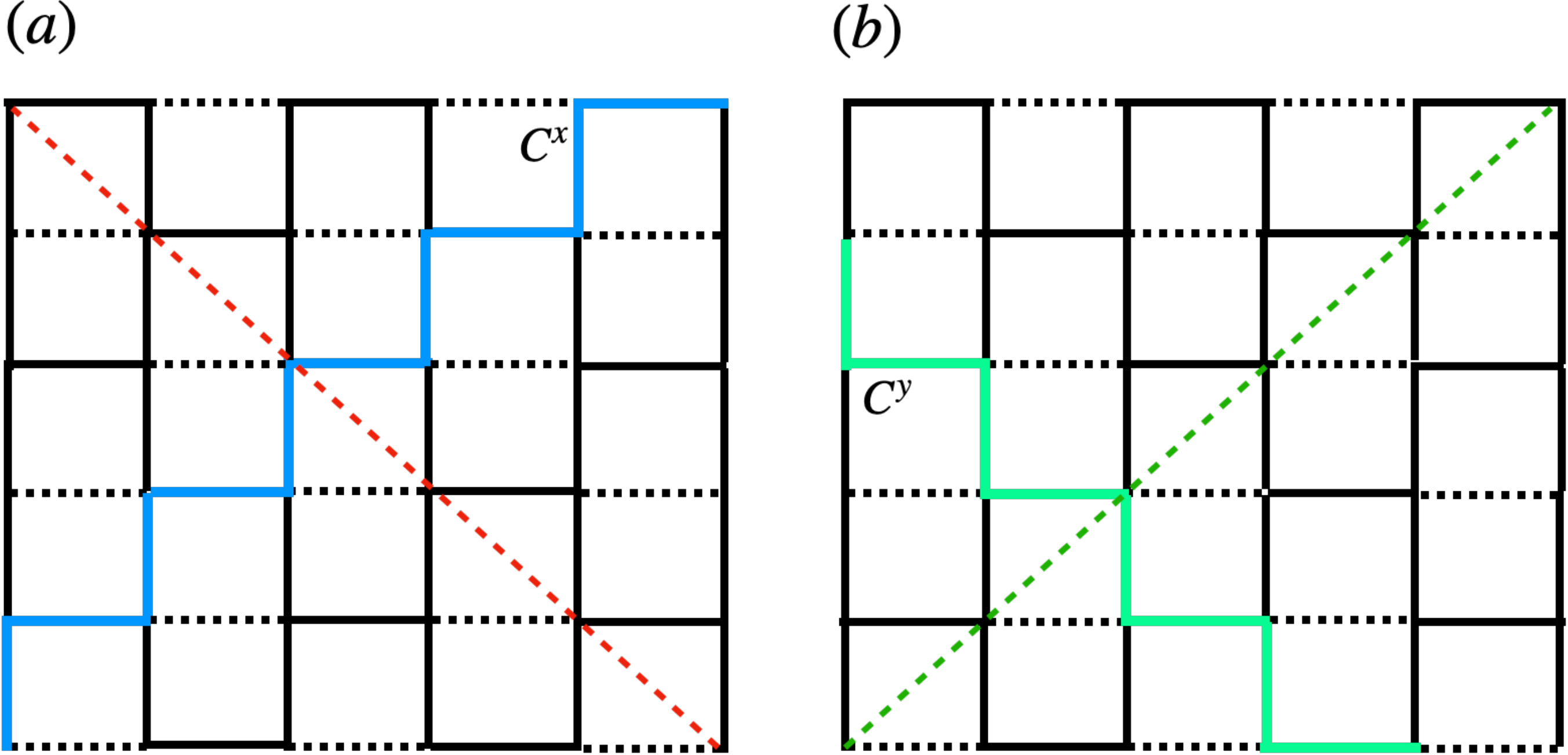}
%{wilsonloops.pdf}
\caption{Two non-contractible loops $C^{x,y}$ for the lattice on a torus. Solid black lines represent non-trivial $\rho^{\gamma}_{ij}=1$ bonds while the dashed black lines are identity bonds. $C^{x/y}$ only include non-trivial bonds. Dashed red and green lines indicates paths used in constructing the Wilson loop operators in eq.~\ref{Eq:Wlsn_1}}.
\label{Fig:S5}
\end{suppfigure}

\noindent for both sublattices. $N_{2/1}$ is the number of unit cells along the two directions implied by $C^{y/x}$. The $\eta^{x/y}$ components obey anti-periodic BC (ABC), while the $z$-components are subject to PBC. 

\subsection{$h_{x} >0, h_{z}=0$}

It is convenient to apply a global rotation $\eta^{x}_{i}  \rightarrow  \tilde{\eta}^{z}_{i}, \eta^{y}_{i}  \rightarrow  \tilde{\eta}^{y}_{i}, \eta^{z}_{i} \rightarrow  -\tilde{\eta}^{x}_{i}$ which maps $H_{g_{2}}$ to

\noindent \begin{align}
    \tilde{H}_{g_2}= & h_{x}  \sum_{i} (-1)^{i_{x}+i_{y}}  \tilde{\eta}^{z}_{i} +
    g_{2} \sum_{\langle ij \rangle_{\gamma} }  \left[ \tilde{\eta}^z_i \tilde{\eta}^z_j \rho^\gamma_{ij} + \tilde{\eta}_i^x \tilde{\eta}_j^x
+  \tilde{\eta}^y_i \tilde{\eta}^y_j \rho^\gamma_{ij} \right],    
\label{Eq:Hg2_hx}
\end{align}

\noindent For a uniform $\rho^{\gamma}_{ij}=1$ configuration, this reduces to a 2D AFM Heisenberg model which has finite staggered pseudo-spins in the GS for any $h_{x} \neq  0$. A non-contractible loop implies ABC for $\tilde{\eta}^{z}_{i} = - \tilde{\eta}^{z}_{i+ N_{1/2}}$. At a semi-classical level, the presence of the staggered field $h_{x}$ implies that such a twist costs an amount of energy which scales with $N_{2/1}$. Moreover, this energy cost diverges for an infinite-size system. This also holds in the subsequent $h_{x} \rightarrow 0$ limit. 

The same conclusion can be reached using a perturbative analysis. It is convenient to re-write 

\noindent \begin{align}
    \tilde{H}_{g_2}= & H_{0} +
    g_{2} \sum_{\langle ij \rangle_{\gamma} } V^{\gamma}_{z, ij} + g_{2} \sum_{\langle ij \rangle_{\gamma} } V^{\gamma}_{\perp, ij},    
\end{align} 

\noindent where

\noindent \begin{align}
H_{0} = & h_{x}  \sum_{i} (-1)^{i_{x}+i_{y}}  \tilde{\eta}^{z}_{i} \\
V^{\gamma}_{z, ij} = &   \tilde{\eta}^z_i \tilde{\eta}^z_j \rho^\gamma_{ij} \\
V^{\gamma}_{\perp, ij} = &   \tilde{\eta}_i^x \tilde{\eta}_j^x + \tilde{\eta}^y_i \tilde{\eta}^y_j \rho^\gamma_{ij}.
\end{align}

\noindent When considering matrix elements, we shall label the expectation value in the GS of $\tilde{H}_{0}$ as $\braket{}_{00}$.  $\braket{}_{01}$ indicates matrix elements between GS and an excited state with a NN pair of flipped pseudo-spins, and similarly $\braket{}_{12}$ for the matrix elements between two excited states with 1 and 2 pairs of flipped pseudo-spins, respectively. The GS has staggered pseudo-spins to zeroth order in $g_{2}$. To first order, the correction to the GS energy is 

\noindent \begin{align}
E^{(1)}_{n, \rho} = & g_{2} \bigg \langle \sum_{\braket{ij}_{\gamma}} V^{\gamma}_{z, ij} \bigg \rangle_{00} \\
= & -   g_{2} \times n \times \rho + \text{Background},
\end{align}

\noindent where we explicitly indicated the number of contributing configurations and their weights. These are due to the Ising exchange terms along the (rotated) $z$ axis, and amount to taking into account all of the bonds in $\tilde{H}_{\rm{String}}$ and $\tilde{H}_{\rm{Coupling}}$. $\rho$ stands for the two cases with either all bonds positive or negative on the string. The  remaining contributions from $\tilde{H}_{\rm{Background}}$ are independent of $\rho$. The corrections to second order are due  exclusively to the (rotated) in-plane terms, which introduce pseudo-spin flips for NNs. Importantly, each flip carries a $(1+ \rho^{\gamma}_{ij})$ factor as 

\noindent \begin{align}
E^{(2)}_{n, \rho} = & g^{2}_{2} \frac{ \sum_{\braket{ij}_{\gamma}} \braket{V^{\gamma}_{\perp, ij}}^{2}_{01}}{\Delta E_{01}}
\\
%= &  - g^{2}_{2}  \frac{\big \langle \sum_{\braket{ij}}( 1+\rho^{\gamma}_{ij})^{2} \big \rangle _{\rm{GS}}}{32h_{x}} \notag \\
 = &  - \frac{g^{2}_{2}}{32 h_{x}} \times  n \times ( 1+\rho)^{2} +  \text{Background},
\end{align}

\noindent where $\Delta E_{01} = -2h_{x}$ is the energy cost of an excited state with a single pair of flipped pseudo-spins. Also note that $\braket{V^{\gamma}_{\perp, ij}} = 2^{-2} (1+ \rho^{\gamma}_{ij})$. To third order, we obtain 

\noindent \begin{align}
E^{(3)}_{n, \rho} = &  g^{3}_{2} \frac{\sum_{\braket{ij}_{\gamma}} \braket{ V^{\gamma}_{\perp, ij}}_{01} 
\left( \braket{\sum_{\braket{ab}_{\gamma}} V^{\gamma}_{z, ab}}_{11} - \braket{\sum_{\braket{ab}_{\gamma}} V^{\gamma}_{z, ab}}_{00}\right) \braket{ V^{\gamma}_{\perp, ij}}_{10}}{(\Delta E_{01})^{2}} 
\notag \\
= & \frac{ g^{3}_{2}}{64 h^{2}_{x}} \left[ n \times (1+\rho)^{2}(8+4\rho) + 2n \times 4(8+4\rho) \right] + \text{Background}.
\end{align}

\noindent These corrections are due to the energy cost introduced by $V^{\gamma}_{z}$ terms in the presence of a pair of flipped pseudo-spins in the string and connecting bonds. To fourth order, the corrections read

\noindent \begin{align}
E^{(4)}_{n, \rho} = & g^{4}_{2} \Bigg( \frac{\sum_{p} \braket{V^{\gamma}_{\perp, ij}}_{01} \braket{V^{\gamma}_{\perp, jk}}_{11} \braket{V^{\gamma}_{\perp, kn}}_{11}\braket{V^{\gamma}_{\perp, ni}}_{10} 
}{(\Delta E_{01})^{3}} 
 + \frac{\sum_{p} \braket{V^{\gamma}_{\perp, ij}}_{01} \braket{V^{\gamma}_{\perp, nk}}_{12} \braket{V^{\gamma}_{\perp, kj}}_{21}\braket{V^{\gamma}_{\perp, ni}}_{10}
}{(\Delta E_{01})^{2} \Delta E_{02}} 
+ (\text{Permut.}/p \leftrightarrow p')
\notag \\
& + \frac{\sum_{\braket{ij}_{\gamma}} \sum_{\braket{jk}_{\gamma}, k\neq i}  \braket{V^{\gamma}_{\perp, ij}}_{01} \braket{V^{\gamma}_{\perp, jk}}_{11}\left( \braket{V^{\gamma}_{\perp, jk}}_{11}\braket{V^{\gamma}_{\perp, ij}}_{10} 
+ \braket{V^{\gamma}_{\perp, ij}}_{11}\braket{V^{\gamma}_{\perp, jk}}_{10} \right)}{(\Delta E_{01})^{3}} 
\notag \\
& + \frac{\sum_{\braket{ij}_{\gamma}} \sum_{ \braket{ab}_{\gamma}, ab \neq ij} \braket{V^{\gamma}_{\perp, ij}}_{01} \braket{V^{\gamma}_{\perp, ab}}_{12}
\left( \braket{V^{\gamma}_{\perp, ab}}_{21}\braket{V^{\gamma}_{\perp, ij}}_{10}
+ \braket{V^{\gamma}_{\perp, ij}}_{21}\braket{V^{\gamma}_{\perp, ab}}_{10}
\right)}{(\Delta E_{01})^{2}\Delta E_{02}}   \notag \\
& - \frac{\sum_{\braket{ij}_{\gamma}}\sum_{\braket{jk}_{\gamma}, k \neq i} \braket{V^{\gamma}_{\perp, ij}}^{2}_{01} \braket{V^{\gamma}_{\perp, jk}}^{2}_{01} }{(\Delta E_{01})^{3}}
- \frac{\sum_{\braket{ij}_{\gamma}}\sum_{\braket{ab}_{\gamma}} \braket{V^{\gamma}_{\perp, ij}}^{2}_{01} \braket{V^{\gamma}_{\perp, ab}}^{2}_{01} }{(\Delta E_{01})^{3}} 
\notag \\
& +  \frac{\sum_{\braket{ij}_{\gamma}}  \braket{V^{\gamma}_{\perp, ij}}_{01}  \left( \braket{ \sum_{ \braket{ab}_{\gamma}} V^{\gamma}_{z, ab}}_{11} - \braket{ \sum_{ \braket{ab}_{\gamma}} V^{\gamma}_{z, ab}}_{11}\right)^{2} \braket{V^{\gamma}_{\perp, ij}}_{10}}{(\Delta E_{01})^{3}} \Bigg)
\end{align}

\noindent The first line includes all pseudo-spin flips around $p/p'$ plaquettes. The next three lines amount to flips on single bonds and on pairs of connected bonds, while the last line contains corrections similar to those at third order. Their effective contribution is 

\noindent \begin{align}
E^{(4)}_{n, \rho} = & -\frac{g^{4}_{2}}{2048 h^{3}_{x}} \Bigg( 20 \times  n\times 4(1+\rho)^{2} +2 \times n \times (1+\rho)^{4}
+  2 \times 4n \times 4(1+\rho)^{2}
+ 2 \times 2n \times 16 \notag \\
& + n \times (1+\rho)^{2}(8+4\rho)^{2} + 2n \times 4(8+4\rho)^{2} - n \times (1+\rho)^{4}
- 2n \times 16 \Bigg) + \text{Background}.
\end{align}

\noindent The energy cost of the string to fourth order is

\noindent \begin{align}
E_{n, \rho=-1} - E_{n, \rho=1} =  n \left( 2g_{2} +  \frac{g^{2}_{2}}{8h_{x}} - \frac{7 g^{3}_{2}}{4h^{2}_{x}} + \frac{129 g^{4}_{2}}{128 h^{3}_{x}} \right) + O(g^{5}_{2}).
\end{align}

\noindent It is apparent that this diverges  as the system size, and thus $n$, grows to infinity. Note that this also implies that the cost of an open string of infinite extent likewise diverges. An open string of length $(n-2)$ is obtained from the non-contractible loop by changing the sign of two neighboring bonds. This is a local perturbation, which costs a finite amount of energy. The energy of the string therefore diverges as that of the parent state. As discussed in the main text, this implies that the visons are confined.  

\subsection{$h_{z} > 0, h_{x}=0$}

In contrast to the case with $h_{x}> 0, h_{z}=0$, here the BCs associated with the loop  do not imply a finite energy cost in a semi-classical approximation. This is due to the fact that the pseudo-spins are staggered along $z$, and $\eta^{z}$ does not couple to the bonds. We further illustrate by considering the Holstein-Primakoff representation for the pseudo-spins~\cite{Auerbach}.

\noindent \begin{align}
\eta^{+}_{i} = & \left(\sqrt{2\eta - n_{b,i}} \right)b_{i} \\
\eta^{-}_{i} = & b^{\dag}_{i}\left(\sqrt{2\eta - n_{b,i}} \right) \\
\eta^{z}_{i} = & \eta - n_{b,i}.
\end{align}

\noindent Here, $b_{i}, b^{\dag}_{i}$ are bosons, $n_{b, i}=b^{\dag}_{i} b_{i}$, and $\eta \gg n_{b, i}$ is the size of the pseudo-spin. The BCs for the pseudo-spins can be implemented by imposing ABCs for the bosons as $b_{i} = - b_{i+N_{2}}$. The remaining analysis follows the usual semi-classical approximation. In particular, as $h_{z} \rightarrow 0$, the zero-point energy of the GS involves a summation over the reciprocal unit cell as~\cite{Auerbach}.

\noindent \begin{align}
E' = 2 \sum_{\mathbf{k}} g_{2} \eta \left( \sqrt{1- \alpha^{2}_{\mathbf{k}}} -1 \right),
\end{align}

\noindent where $\alpha_{\mathbf{k}}$ is the usual form factor for a square lattice. ABC amount to a trivial translation of the entire reciprocal unit cell in the extended BZ by $\mathbf{G}_{1/2}/2N_{1/2}$, where $\mathbf{G}_{2}$ is a reciprocal lattice vector. The correction to the classical GS energy is invariant. Consequently, at this level of approximation, the string is degenerate with the uniform $\rho^{\gamma}_{ij}=1$ state. 

We reach the same conclusion using a perturbative analysis. In contrast to the $h_{x}>0$ case, the first-order corrections are independent of $\rho^{\gamma}_{ij}$:

\noindent \begin{align}
E^{(1)}_{n, \rho} = & -  g_{2} \sum_{\braket{ij}}  \notag \\
= & -n g_{2} + \text{Background},
\end{align}

\noindent since the Ising interaction along $z$ does not couple to the bonds. The second-order contribution,  due to pairs of NN pseudo-spin flips, reads

\noindent \begin{align}
E^{(2)}_{n, \rho} = & -  g^{2}_{2}  \frac{ \sum_{\braket{ij}} (\rho^{\gamma}_{ij})^{2}}{8h_{z}} \notag \\
 = & - n g^{2}_{2}  \frac{1}{8h_{z}} + \text{Background}.
\end{align} 

\noindent Note the important difference w.r.t. the $h_{x} > 0$ case, where each flip of NN pseudo-spins was multiplied by a $(1+\rho^{\gamma}_{ij})$ factor. Here, the corrections are independent of bond configuration. A similar conclusion holds at third order. At fourth order, we obtain the plaquette terms

\noindent \begin{align}
E^{(4)}_{n, \rho} = & - \frac{g^{4}_{2}}{32 h^{3}_{x}} \Bigg[ \sum_{p} \Bigg( \prod_{\braket{ij} \in p/p'} \rho^{\gamma}_{ij} \Bigg) + p \leftrightarrow p'  \Bigg] + \hdots, 
\end{align} 

\noindent where we omitted contributions similar to those at third order, which likewise are independent of the bonds. The terms shown above involve products of $\rho$'s around $p/p'$ plaquettes. Since each plaquette contains at least two bonds on the string, these are also trivial. However, similar corrections involving pseudo-spin flips around closed loops, which intersect the string an odd number of times, are non-trivial. It is straightforward to check that only non-contractible loops, which wrap around the torus at least once, can contribute. These occur at leading $n$th order and scale with a factor of $g^{n}_{2}/h^{n-1}_{z}$. Consequently, each is exponentially suppressed as the system size, and therefore $n$, goes to infinity. The string becomes degenerate with the uniform $\rho^{\gamma}_{ij}=1$ state in this limit. This also implies an upper bound on the energy cost of any open string of length $n-2$. The latter can be obtained by changing the sign of any two NN bonds, which is a local perturbation, implying a finite energy cost. This also holds as the system size goes to infinity. As discussed in the main text, it follows that visons are deconfined. 

\section{Connection between Hubbard and $\eta-\rho$ models}

In this section, we show that the 
GSs of the Hubbard model and $H_{\eta-\rho}$ models are equivalent. 

We start with the Hubbard model (eq.~3 in the main text). In general, the GS for large $J$ is characterized by a finite expectation value of the gauge-invariant correlators 

\begin{eqnarray}
\mathcal{C}^{x}_{ij} = \langle N^{x}_{i} \prod_{i'j' \in (i,j)} u_{1 i' j'}u_{2 i' j'} N^{x}_{j} \rangle
\end{eqnarray}
where
\begin{align}
    N_{i/j}^x =(-1)^{i_{x}+i_{y}}\expval{f^\dagger_{2 i}f_{1 i}+f^\dagger_{1 i}f_{2 i}}
\end{align}
are staggered, pseudo-spins along $x$, which are connected via a gauge string, and similarly for $y$ components. We write the correlator in terms of Majorana fermions as
\begin{eqnarray}
\mathcal{C}^{x}_{ij}
=\prod_{i'j' \in (i,j)} (-1)^{i_{x}+i_{y}+i'_{x}+i'_{y}}\langle (c_{1i}^xc_{2i}^x+c_{1i}^yc_{2i}^y)  u_{1 i' j'}u_{2 i' j'}(c_{1j}^xc_{2j}^x+c_{1j}^yc_{2j}^y)\rangle
\end{eqnarray}

We next consider a similar correlator in the $\pi$-flux GS of $H_{\eta-\rho}$, (eq. 19 in the main text)

\noindent \begin{align}
\mathcal{D}^{x}_{ij}= &  (-1)^{i_{x}+i_{y} + i_{x'}+ i_{y'}} \Bigg \langle   \eta^{x}_{i} \left( \prod_{i'j' \in C_{ij}} \rho^{\gamma}_{ij} \right) \eta^{x}_{j} \Bigg \rangle.
\label{Eq:D}
\end{align}

\noindent The correlator commutes with $W_{\nu p/p'}$. It follows from the form of the GS of $H_{\eta-\rho}$ that $\mathcal{D}^{x}_{ij}$ is finite for $h_{x}>0$ and any $i,j$. We next perform a rotation on all of the spin ($\sigma$) DOF of the second layer as

\begin{align}
\eta^{x}_{i} \rightarrow \tilde{\eta}^x_{i}&= \frac{1}{4}(\sigma_{1i}^x\sigma_{2i}^x+\sigma_{1i}^y\sigma_{2i}^y) \nonumber \\
\eta^{y}_{i} \rightarrow \tilde{\eta}^y_{i}&= \frac{1}{4}(\sigma_{1i}^y\sigma_{2i}^x-\sigma_{1i}^x\sigma_{2i}^y) \nonumber
\\
\eta^{z}_{i} \rightarrow \tilde{\eta}^z_i&= \frac{1}{4}(\sigma_{1i}^z-\sigma_{2i}^z). 
\end{align}

\noindent  We introduce a Majorana representation for the rotated pseudo-spins as

\noindent \begin{align}
\tilde{\eta}^x_{i}=(-1)^{i_{x}+i_{y}}b_{1i}^4b_{2i}^4(c_{1i}^xc_{2i}^x+c_{1i}^yc_{2i}^y)),
\end{align}

\noindent as well as for the string 

\noindent \begin{align}
\prod_{i'j' \in C_{ij}} \rho^{\gamma}_{ij} =b_{1 i}^4 b_{2 i}^4 c_{1 i}^x c_{2 i}^y c_{1 i}^x c_{2 i}^y \prod_{i'j' \in (i,j)}u_{1 i'j'}u_{2 i'j'}b_{1 j}^4b_{2 j}^4 c_{1 j}^xc_{2 j}^yc_{1 j}^xc_{2 j}^y
\end{align}

\noindent where we used $\rho^{\gamma}_{ij} =\prod_{\nu} \tau^{\gamma}_{\nu i} \tau^{\gamma}_{\nu j}$ and  $\tau_{\nu i}^\gamma=b_{\nu i}^4b_{\nu i}^\gamma c_{\nu i}^xc_{\nu i}^y$. Substituting these in eq.~\ref{Eq:D}, we obtain

\noindent \begin{align}
\mathcal{D}^{x}_{ij}=\prod_{i'j' \in (i,j)}(-1)^{i_{x}+i_{y}+i'_{x}+i'_{y}}\langle (c_{1i}^xc_{2i}^x+c_{1i}^yc_{2i}^y) u_{1 i' j'}u_{2 i' j'}(c_{1j}^xc_{2j}^x+c_{1j}^yc_{2j}^y)\rangle_{\Psi_{\rm{GS}}}.
\end{align}

\noindent This expression is formally identical to $\mathcal{C}^{x}_{ij}$.

\section{Topological degeneracy of the ground state of $H_{\eta-\rho}$}

In this section we show that the GS of $H_{\eta-\rho}$ exhibits $\mathbb{Z}_{2}$ topological order for $h_{x}>0, h_{z}=0$ and $\mathbb{Z}_{2} \times \mathbb{Z}_{2}$ topological order for $h_{z}>0, h_{x}=0$.

\subsection{$\mathbb{Z}_{2}$ topological order}

For $h_{x}> 0, h_{z}=0$, the GS of $H_{\eta-\tp}$ has the form 

\noindent \begin{align}
  \ket{\Psi_{\rm{GS}}} = & \prod_{\nu, p} \frac{1}{4}\left(1-W_{\nu p}  \right) \left(1-W_{\nu p'}  \right) \ket{\tilde{\eta}_{xx}; \phi_{0}}  + O\left(\frac{h_{x}}{g_{4}} \right). 
  \label{Eq:Prjc}
\end{align}

\noindent where $\ket{\tilde{\eta}_{xx}; \phi_{0}}$ is the GS of the 2D Heisenberg model for finite $h_{x}$, while $\phi_{0}$ is an orbital configuration with all $q^{\gamma}_{i}=-1$, which implies that it has uniform $\rho^{\gamma}_{ij}=1$ bonds. Small corrections to this state occur because $h_{x}$ does not commute with the flux operators $W_{\nu p/p'}$. 

The projection of $\ket{\tilde{\eta}_{xx}; \phi_{0}}$  onto the $\pi$-flux state in eq.~\ref{Eq:Prjc} contains factors of the form $W_{\nu p} W_{\nu p'}$, where $p/p'$ are in the same unit cell. When operating on $\ket{\tilde{\eta}_{xx}; \phi_{0}}$ these preserve all pseudo-spin and $\rho^{\gamma}_{ij}$ bonds. However, they change the sign of all pairs of NN pairs of $q^{\gamma}$'s around the unit cell. To illustrate, the NN pair $q^{z}_{1}=-1, q^{z}_{j}=-1$ is mapped onto $q^{z}_{1}=1, q^{z}_{j}=1$, with $\rho^{\gamma}_{ij}=1$ in both cases, and similarly for the remaining NN pairs. However, since these states are obtained from the $\phi_{0}$ configuration via local changes in the set of $q$'s, they are not topologically distinct. The $W_{\nu p} W_{\nu p'}$ operators are in this sense the analogs of plaquette operators in the toric code, while the signs of NN $q^{\gamma}$'s can be used to define effective bonds. 

We construct topologically distinct configurations by operating with 

\noindent \begin{align}
F^{x/y} = &\prod_{i\in C^{x/y}}\tau^{x/y}_{1i}.
\end{align}

\noindent Here, the orbital operators $\tau^{x/y}_{1i}$ are applied on every site along non-contractible loops $C^{x/y}$, as illustrated in Supplementary figure~\ref{Fig:S5}.  $F^{x/y}$  commutes with both $H_{\eta-\rho}$ and $W_{\nu, p/p'}$, and has the effect of changing the signs of all NN $q$'s along the loop. Note that $\tau^{x/y}_{2i}$ would have the same effect. Any state thus obtained is degenerate with $\ket{\Psi_{\rm{GS}}}$. 

In order to capture the topology of the GS manifold, we introduce the Wilson loop operators 

\noindent \begin{align}
  I^{x/y}=&\prod^{'}_{ i,j \in C^{y/x}} q_i^{y/x} q_j^z,
  \label{Eq:Wlsn_1}
\end{align}

\noindent where $\prod^{'}$ denotes every other site on $C^{y/x}$ (Supplementary figure~\ref{Fig:S5}). Since $\left[I^{x/y}, H_{\eta-\rho} \right]=0$ and $\left[I^{x/y}, W_{\nu p/p'} \right]=0$, $\ket{\Psi_{\rm{GS}}}$ can be labeled by $\lambda_{x}= \lambda_{y}=1$. 
More generally, the states 

\noindent \begin{align}
\ket{\Psi_{\rm{GS}}; \lambda_{x}, \lambda_{y}} = \left( F^{x} \right)^{\frac{1+\lambda_{x}}{2}} \left( F^{y} \right)^{\frac{1+\lambda_{y}}{2}} \ket{\Psi_{\rm{GS}}}.
\end{align}

\noindent belong to four distinct topological sectors. The four-fold degeneracy is characteristic of $\mathbb{Z}_{2}$ topological order. This procedure exhausts the degeneracy of the GS manifold, since any non-local changes in the pseudo-spin configurations cost finite energy for  $h_{x} > 0$. Similarly, changes in the $\rho^{\gamma}$ bonds are equivalent to the strings discussed in Supplementary section~\ref{Sec:Strn} and are likewise gapped for $h_{x}> 0$. 

\subsection{$\mathbb{Z}_{2} \times \mathbb{Z}_{2}$ topological order}

The previous arguments for $h_{x}> 0, h_{z}=0$ also hold for $h_{z} > 0, h_{x}=0$. The $\mathbb{Z}_{2}$ degeneracy due to configurations of the local orbital states $q^{\gamma}$ consistent with a given set of $\rho^{\gamma}$ bonds still holds. However, as shown in Supplementary section~\ref{Sec:Strn}, non-contractible loops of flipped ($\rho^{\gamma}_{ij} \rightarrow - \rho^{\gamma}_{ij}$) bonds becomes degenerate with $\ket{\Psi_{\rm{GS}}}$ in the infinite-size limit. In order to account for this additional degeneracy, we can define the Wilson loop operators 

\noindent \begin{align}
I^{x/y}_{A} = \prod_{i,j \in C^{y/x}} \rho^{\gamma}_{ij},
\end{align}

\noindent with eigenvalues $\lambda_{A,x/y}= \pm 1$. These commute with $H_{\eta-\rho}$ and $W_{\nu, p/p'}$. For any eigenvalues of $I_{x/y}$ (eq.~\ref{Eq:Wlsn_1}), we can therefore distinguish four additional topologically distinct states 

\noindent \begin{align}
\ket{\Psi_{\rm{GS}}; \lambda_{x}, \lambda_{y}, \lambda_{A,x}, \lambda_{A, y}} = \left( L^{x} \right)^{\frac{1+\lambda_{A, x}}{2}} \left( L^{y} \right)^{\frac{1+\lambda_{A, y}}{2}} \ket{\Psi_{\rm{GS}}, ; \lambda_{x}, \lambda_{y}}. 
\end{align}   

\noindent The operators

\noindent \begin{align}
L^{x/y} = \prod^{'}_{i \in C^{x/y}} \tau^{x/y}_{i}
\end{align}

\noindent act along every other site on path $C^{x/y}$ to create non-contractible loops of flipped $\rho^{\gamma}$ bonds. The GS thus exhibits a sixteen-fold degeneracy on the torus, as consistent with a $\mathbb{Z}_{2} \times \mathbb{Z}_{2}$ topological order.

\bibliography{references}